\documentclass[12pt]{article}
\usepackage{a4wide}
\usepackage{amssymb}
\usepackage{graphicx}
\begin{document}
{\renewcommand{\thefootnote}{\fnsymbol{footnote}}
\hfill  IGC--09/11--4\\
\medskip
\begin{center}
{\LARGE  Effective Constraints and Physical Coherent States\\ in Quantum Cosmology: A Numerical Comparison}\\
\vspace{1.5em}
Martin Bojowald\footnote{e-mail address: {\tt bojowald@gravity.psu.edu}}
and Artur Tsobanjan\footnote{e-mail address: {\tt axt236@psu.edu}}
\\
\vspace{0.5em}
Institute for Gravitation and the Cosmos,\\
The Pennsylvania State
University,\\
104 Davey Lab, University Park, PA 16802, USA\\
\vspace{1.5em}
\end{center}
}

\setcounter{footnote}{0}

\newcommand{\lP}{\ell_{\mathrm P}}

\newcommand{\md}{{\mathrm{d}}}
\newcommand{\tr}{\mathop{\mathrm{tr}}}
\newcommand{\sgn}{\mathop{\mathrm{sgn}}}

\begin{abstract}
 A cosmological model with a cyclic interpretation is introduced,
 which is subject to quantum back-reaction and yet can be treated
 rather completely by physical coherent state as well as effective
 constraint techniques. By this comparison, the role of quantum
 back-reaction in quantum cosmology is unambiguously
 demonstrated. Also the complementary nature of strengths and
 weaknesses of the two procedures is illustrated. Finally, effective
 constraint techniques are applied to a more realistic model filled
 with radiation, where physical coherent states are not available.
\end{abstract}

\section{Introduction}

To understand the generic behavior of a quantum system, effective
equations\footnote{We use the term ``effective equations'' in a
general sense to encompass equations that might follow from an
effective action, or be derived in a canonical way. Systematic
procedures exist for both ways, and the results agree in cases where
they have been explicitly applied \cite{EffAcQM,EffAc}.} are useful. In
contrast to individual wave functions, or even just stationary states,
they directly provide (approximate) equations for time-dependent
expectation values. Since the dynamics of expectation values depends
on the whole motion of a state --- by quantum back-reaction, all
moments of a state couple to the expectation values --- these
equations in general differ from the classical ones by quantum
corrections. If quantum corrections, for instance an effective
potential, can be found explicitly, an interpretation of quantum
dynamics in generic terms becomes much easier. Such results are more
general compared to conclusions based on individual states.

Especially in quantum cosmology, the ability to draw generic
conclusions is important. Not much is known about the state of the
universe except, perhaps, that it currently can well be considered
semiclassical. But semiclassicality is not a sharp notion, and so wide
classes of states, differing for instance by the sizes of their
quantum fluctuations or correlations, are still allowed. In any such
situation, a generic analysis is called for, most crucially when
long-term evolution is involved, or when one evolves toward
strong-curvature regimes such as the big bang singularity where
quantum effects of all kinds are expected to be important.

It is sometimes suggested, at least implicitly (and especially in the
context of loop quantization), that quantum cosmology might somehow be
different from other quantum systems, and that quantum back-reaction
could be ignored in its effective equations. Quantum back-reaction
might be weak for certain states or in certain regimes, especially for
models close to solvable ones, but this observation cannot be
generalized. Like the harmonic oscillator in quantum mechanics, there
are harmonic cosmologies \cite{BouncePert,BounceCohStates} where
expectation values of states follow exactly the trajectories of a
corresponding classical system. Such systems are entirely free of
quantum back-reaction. For ``small anharmonicity'', quantum
back-reaction might still be weak, as it is realized in quantum
cosmology for matter dominated by its kinetic energy density
\cite{QuantumBounce}. But the tough reality of stronger deviations
from the solvable ideal of harmonic systems can introduce severe
quantum back-reaction, which must be studied in an unbiased and
systematic way.

Here, we introduce a model of quantum cosmology which is not solvable
but still treatable by two rather different methods: effective
constraints and physical states. The model is an anisotropic cosmology
of locally rotationally symmetric (LRS) Bianchi I symmetry type,
filled with an isotropic, slightly sub-stiff fluid of negative energy
density $\rho(a)\propto -\log(a)^2/a^6$ where $a$ is the average scale
factor of the anisotropic geometry. With this specific density, the
model becomes treatable by physical coherent states, which justifies
its contrived and exotic form. Effective constraint techniques
\cite{EffCons,EffConsRel}, applicable much more widely, do not require
such a tailored matter source; they are considerably more powerful. As
we will see by the explicit comparison of this paper, they capture the
information in semiclassical physical states to an excellent
degree. Moreover, effective techniques self-consistently determine
their ranges of validity. The general applicability of effective
constraints will be demonstrated by our final analysis of a model
whose matter content, more realistically, is pure radiation.

\section{The model}
\label{sec:model}

An anisotropic Bianchi I model has a line element
\[
 \md s^2=-\md t^2+\sum_{I=1}^3 a_I(t)^2 (\md x^I)^2
\]
with three independent scale factors $a_I(t)$ as functions of proper
time $t$. For a canonical formulation, we denote the momenta of
$h_I:=a_I^2$ by $\pi^J$. It is convenient to introduce Misner
variables $(\alpha,\beta_+,\beta_-)$ and their momenta
$(p_{\alpha},p_+,p_-)$ by the canonical transformation \cite{Mixmaster}
\begin{eqnarray}
 \frac{1}{2}\log (h_1/a_0^2)=: \alpha+\beta_++\sqrt{3}\beta_-\quad&,&\quad
 2h_1\pi^1 =:
 \frac{1}{3}p_{\alpha}+\frac{1}{6}p_++\frac{1}{2\sqrt{3}}p_-\\
  \frac{1}{2}\log (h_2/a_0^2)=: \alpha+\beta_+-\sqrt{3}\beta_-\quad&,&\quad
 2h_2\pi^2 =:
 \frac{1}{3}p_{\alpha}+\frac{1}{6}p_+-\frac{1}{2\sqrt{3}}p_-\\
 \frac{1}{2}\log (h_3/a_0^2) =: \alpha-2\beta_+ \quad&,&\quad
 2h_3\pi^3=:
 \frac{1}{3}p_{\alpha}- \frac{1}{3}p_+\,.
\end{eqnarray}
In these definitions, $a_0$ is a reference scale factor (e.g.\
$a_0=\sqrt{h_1h_2h_3}$ at one moment of time) introduced to be
insensitive to coordinate rescalings.  

Canonical dynamics in general relativity is determined by the
Hamiltonian constraint
\begin{equation}\label{Cgravhom}
 C_{\rm grav}=
   \frac{\pi_{ab}\pi^{ab}-\frac{1}{2}(\pi^a{}_a)^2}{\sqrt{\det h}}-
\frac{\sqrt{\det h}}{(16\pi
   G)^2} {}^{(3)}R=0
\end{equation}
with the spatial metric $h_{ab}$, its Ricci scalar ${}^{(3)}R$ and
its momenta $\pi^{ab}$.  Reduced to Bianchi I metrics in Misner
variables, specified to a lapse function $N=\sqrt{\det h}=
a_1a_2a_3$, it simplifies considerably to the form
\begin{equation} \label{VacConstr}
 C_{\rm Bianchi\: I}=\frac{1}{24} (-p_{\alpha}^2+p_+^2+p_-^2)\,.
\end{equation}
As one can check directly, the constraint generates the correct
Hamiltonian equations of motion in coordinate time, from which the Kasner
solutions follow.

We now restrict the model further by requiring the anisotropy
parameter $\beta_-$ and its momentum $p_-$ to vanish:
$\beta_-=0=p_-$. In this way, which can easily be confirmed to be
consistent with the equations of motion, we enhance the symmetry and
leave two independent gravitational variables: the logarithm of the
average scale factor, $\alpha$, and one anisotropy parameter
$\beta_+$. The resulting Hamiltonian constraint is equivalent to that
of a free, massless relativistic particle.

To introduce a ``potential'', we will work with a matter source whose
energy density is
\begin{equation}
 \rho = - \rho_0\frac{\log(a/a_0)^2}{(a/a_0)^6}= - \rho_0 a_0^6 e^{-6\alpha}\alpha^2\,
\end{equation}
where, in addition to $a_0$ already introduced, $\rho_0$ is the matter
energy density at some ``initial'' time. In the presence of matter,
its density multiplied with $a^6$ is to be added to the constraint
(\ref{VacConstr}). It becomes\footnote{If we introduce a finite region
of coordinate size $V_0$ to integrate out homogeneous quantities such
as the symplectic form $\int\md^3x \delta h_{ab}\wedge\delta\pi^{ab}=
V_0\delta h_I\wedge\delta \tilde{\pi}^I= \delta h_I\wedge\delta\pi^I$,
the momenta $\pi^I=V_0\tilde{\pi}^I$ depend on the rather arbitrary
$V_0$. The lapse function as chosen here would then be homogeneous in
$V_0$, too, such that $V_0^2a^6\rho$ is the matter contribution to the
constraint. With this, all its terms scale in the same way if $V_0$ is
changed. Solving the constrained system, we obtain the same reduced
phase space for all choices of $V_0$.}
\begin{equation}
 C_{\rm LRS}=-p_{\alpha}^2+p_+^2 - \alpha^2 \nu
\end{equation}
if we make convenient (and irrelevant) choices for the prefactors
and define $\nu = \rho_0 a_0^6 $. Our constraint
then becomes that of a ``relativistic harmonic oscillator'' as
studied in \cite{EffConsRel}:
\begin{equation}
 p_+^2= p_{\alpha}^2+\nu\alpha^2\,.
\end{equation}
In this analogy, we consider $\beta_+$ as our time variable, and
$p_+$ as the corresponding ``energy.'' Evolution of $\alpha$ and
$p_{\alpha}$ with respect to $\beta_+$ is then generated by the
Hamiltonian $p_+=\pm\sqrt{p_{\alpha}^2+\nu\alpha^2}$.

A more realistic matter content could be chosen as radiation, with
an energy density $\rho\propto a^{-4}=e^{-4\alpha}$. Here, the
Hamiltonian is $p_+=\pm\sqrt{p_{\alpha}^2-\nu e^{2\alpha}}$.

\subsection{Classical behavior}
\label{sec:classical}

We define $H=\sqrt{p_{\alpha}^2+\nu\alpha^2}$, such that our
constraint solved for $p_+$ takes the deparameterized form
\[
 C_{\rm deparametrized}= p_+\pm H(\alpha,p_{\alpha})=0\,.
\]
(From now on we choose $p_+$ positive, to be specific.) This
constraint generates Hamiltonian equations of motion for the
anisotropies $\dot{p}_+=0$ and $\dot{\beta}_+=1$, allowing us to
identify the time parameter along its flow with $\beta_+$ as an
internal time. All derivatives in Hamiltonian equations of motion are
then with respect to $\beta_+$, specifically
\[
 \frac{\md\alpha}{\md\beta_+}= \{\alpha,H\}= \frac{p_{\alpha}}{H}\quad,\quad
 \frac{\md p_{\alpha}}{\md\beta_+}= \{p_{\alpha},H\}= -\frac{\nu\alpha}{H}\,.
\]
Since $H=\mp p_+$ is constant in $\beta_+$, we can combine these
equations to a second order one for $\alpha(\beta_+)$,
$\md^2\alpha/\md\beta_+^2= -\nu\alpha/H^2$ solved by
\begin{equation}
 \alpha(\beta_+)= A\sin(\beta_+\sqrt{\nu}/H)+B\cos(\beta_+\sqrt{\nu}/H)
\end{equation}
with integration constants $A$ and $B$.  The equation of motion for
$\alpha_+$ tells us that
\begin{equation}
 p_{\alpha}(\beta_+)= \sqrt{\nu} \left(A\cos(\beta_+\sqrt{\nu}/H)-B\sin(\beta_+\sqrt{\nu}/H)\right)\,.
\end{equation}
With this, the integration constants can be related to $H$ by $H^2=\nu
\left(A^2+B^2\right)$. Solutions in phase space are ellipses of axis
lengths $H=|p_+|$ and $\sqrt{\nu}H$, traversed in time $\beta_+$ with
frequency $(2\pi H/\sqrt{\nu})^{-1}$.

To derive the behavior with respect to proper time, we use the
original constraint $C_{\rm LRS}$ not yet deparameterized, and
remember that we chose a lapse function $N=a_1a_2a_3= a_0^3
e^{3\alpha}$. Thus, the equation of motion for $\beta_+$ with
respect to proper time $\tau$ is
\[
 \frac{\md\beta_+}{\md\tau}= \{\beta_+, e^{-3\alpha} C_{\rm LRS}\}=
 2e^{-3\alpha} p_+a_0^{-3}\,.
\]
From here, we determine proper time as a function of $\beta_+$ by
integrating
\begin{equation}
\tau= \frac{a_0^3}{2p_+} \int e^{3\alpha(\beta_+)}\md\beta_+
\label{eq:proper}
\end{equation}
with our solution $\alpha(\beta_+)$. Inverting the solution allows us
to insert $\beta_+(\tau)$ into $\alpha(\beta_+)$ and
$p_{\alpha}(\beta_+)$, resulting in solutions as functions of proper
time.

It is not easy to integrate $\tau(\beta_+)$ explicitly, but it is
clear from the integrand that $\tau(\beta_+)$ is a monotonic,
one-to-one function which can be inverted globally. Thus,
$\alpha(\tau)$ and $p_{\alpha}(\tau)$ are defined and finite for all
values of proper time. There is no singularity in this model. Clearly,
the negative amount of matter energy violates energy conditions
sufficiently strongly to provide the cyclic bouncing solutions
embodied by ellipses in the $(\alpha,p_{\alpha})$-plane.

One can see this directly from the Friedmann-type equation resulting
from the Hamiltonian constraint. We have $p_{\alpha}^2=
p_+^2-\nu\alpha^2$, where $p_+$ is constant and $p_{\alpha}$ can be
obtained from the equation $\md \alpha/\md\tau= \{\alpha,
a_0^{-3}e^{-3\alpha}C_{\rm LRS}\}= -2a_0^{-3}e^{-3\alpha}
p_{\alpha}$, relating it to the derivative of $\alpha$ by proper
time $\tau$. Thus, the constraint equation reads
\[
 \frac{1}{4}a_0^6e^{6\alpha}
 \left(\left(\frac{\md\alpha}{\md\tau}\right)^2-
 \frac{4}{a_0^6e^{6\alpha}}(p_+^2-\nu\alpha^2)\right)=0
\]
and for $a=a_0\exp(\alpha)$ implies
\begin{equation}
 \left(\frac{\dot{a}}{a}\right)^2= \frac{4p_+^2}{a^6}- 4\rho_0\frac{(\log
 a/a_0)^2}{(a/a_0)^6}\,.
\end{equation}
On the right hand side, we identify $4p_+^2/a^6$ as the anisotropic
shear term, and the contribution $-4\rho_0a_0^6(\log a/a_0)^2/a^6$ as
our exotic energy density.

The Friedmann equation shows when $\dot{a}$ can vanish, which is
realized for $\log a/a_0=\pm p_+/\sqrt{\nu}$, two solutions which
correspond to the maximal and minimal $\alpha$ along our circles. The
extrema indeed give us a maximum for $\log a/a_0= p_+/\sqrt{\nu}$ and
a minimum for $\log a/a_0=-p_+/\sqrt{\nu}$: The sign of $\ddot{a}$, by
the Raychaudhuri equation, is determined by the sign of $\rho+3P$ of
all the matter sources combined. Here, pressure is obtained from the
energy density as the negative derivative of energy by volume, i.e.\
by
\begin{equation}
 P=-\frac{\partial E}{\partial V}=-\frac{ \md(a^3\rho(a))}{3a^2\md a}
\end{equation}
resulting in $P_{\rm neg}= \rho_{\rm neg}(1-\frac{2}{3}(\log
a/a_0)^{-1})$ for the negative energy fluid with $\rho_{\rm neg}=
-4\nu(\log a/a_0)^2/a^6$, and $P_{\rm shear}=\rho_{\rm
shear}$ for the shear contribution with $\rho_{\rm shear}=
4p_+^2/a^6$. For $\rho+3P$, this gives
\[
 \rho+3P= \rho_{\rm neg} \left(4-\frac{2}{\log (a/a_0)}\right)+
 4\rho_{\rm shear}
\]
which at the extrema $\log a/a_0=\pm p_+/\sqrt{\nu}$ evaluates to $\pm
8\sqrt{\nu}p_+/a^6$. The extrema provide a maximum at $\log
a/a_0=p_+/\sqrt{\nu}$ and a minimum at $\log a/a_0=-p_+/\sqrt{\nu}$,
and the evolution is that of a cyclic model with infinitely many
bounces and recollapses.

For the model with radiation, we have equations of motion
$\md\alpha/\md\beta_+= p_{\alpha}/H$ and $\md p_{\alpha}/\md\beta_+=
\nu e^{2\alpha}/H$, with $H$ still constant. The second order
equation for $\alpha(\beta_+)$ becomes $\md^2\alpha/\md\beta_+^2=
\nu e^{2\alpha}/H^2$. For the solution of this model, see
Section~\ref{sec:rad_universe}.

\subsection{Quantum representation}
\label{sec:state}

To represent the model as a quantum system, we start with the
obvious kinematical Hilbert space $L^2({\mathbb
R}^2,\md\alpha\md\beta_+)$ of square integrable wave functions of
two variables, $\beta_+$ and $\alpha$. The momentum operators are
derivatives times $-i\hbar$, as usual. To arrive at observable
information and physical states, we have to implement the constraint
operator\footnote{Henceforth, for this model, we drop $\nu$\ for
simplicity. It can be absorbed into the variables by first
dividing the constraint throughout by $\sqrt{\nu}$, followed by the
canonical transformation $\alpha' = \nu^{\frac{1}{4}}\alpha$,
$p_{\alpha}' = \nu^{-\frac{1}{4}}p_{\alpha}$, $\beta_+'
=\nu^{\frac{1}{4}} \beta_+$, $p_+' = \nu^{-\frac{1}{4}} p_+$.}
\begin{equation}
 \hat{C}= \hat{p}_+^2-\hat{p}_{\alpha}^2-\hat{\alpha}^2=
 -\hbar^2\frac{\partial^2}{\partial^2\beta_+}+
 \hbar^2\frac{\partial^2}{\partial^2\alpha^2}-\alpha^2\,,
\end{equation}
find its kernel and equip it with a physical inner product.

To analyze the quantum constraint, we reformulate it in a first-order
way in our time variable $\beta_+$ by taking a square root:
\begin{equation} \label{eq:schroedinger}
-\frac{\hbar}{i} \frac{\rm d}{{\rm d} \beta_+} \Psi(\alpha, \beta_+)
= \pm\left( \hat{p}_{\alpha}^2 + \hat{\alpha}^2
\right)^{\frac{1}{2}} \Psi(\alpha, \beta_+)=: \pm \hat{H}
\Psi(\alpha, \beta_+)\,,
\end{equation}
introducing the deparametrized Hamiltonian operator $\hat{H}=
(\hat{p}_{\alpha}^2+\hat{\alpha}^2)^{1/2}$.  All solutions of this
equation can be expressed as
\begin{equation} \label{States}
 \Psi_{\pm}(\alpha,\beta_+)= \sum_{n=0}^{\infty} c_n \varphi_n(\alpha)
 \exp(\mp i \lambda_n \beta_+/\hbar)
\end{equation}
where $\varphi_n$ are the eigenstates of $\hat{H}$ with eigenvalues
$\hbar\lambda_n$. Since $\hat{H}$ is the square root of the (positive
definite) harmonic oscillator Hamiltonian (with ``mass'' $m=1/2$ and
``frequency'' $\omega=2$), its eigenstates are of the well-known form,
with eigenvalues $\lambda_n= \sqrt{(2n+1)\hbar}$.

In (\ref{States}), the subscript $\pm$ indicates the sign taken when
solving for $p_+$ by a square root. Solutions naturally split into two
classes, positive-frequency solutions $\Psi_+$ and negative-frequency
solutions $\Psi_-$. The separation into two classes becomes relevant
when we introduce the inner product
\begin{equation}
 (\Psi,\Phi)_{\rm aux}:= i\int\md\alpha
 \left(\Psi^*(\alpha,\beta_+)\frac{\partial}{\partial\beta_+}
 \Phi(\alpha,\beta_+)-
 \Phi(\alpha,\beta_+)\frac{\partial}{\partial\beta_+}
 \Psi^*(\alpha,\beta_+)\right)
\end{equation}
of Klein--Gordon form. Although $\beta_+$ appears on the right hand
side, the inner product evaluated on solutions (\ref{States}) is time
independent. However, $(\cdot,\cdot)_{\rm aux}$ is not positive
definite: it is positive for positive-frequency solutions, but
negative for negative-frequency solutions. A positive-frequency
solution is automatically orthogonal to a solution with the sign of
its frequency flipped. To correct the sign, we define the physical
inner product as
\begin{equation}
 \langle\Psi,\Phi\rangle:= \left\{\begin{array}{cl} (\Psi,\Phi)_{\rm aux} &
 \mbox{if $\Psi$ and $\Phi$ are positive-frequency}\\
-(\Psi,\Phi)_{\rm aux} & \mbox{if $\Psi$ and $\Phi$ are
 negative-frequency}\\
 0 & \mbox{otherwise}\end{array}\right.
\end{equation}
and extend it linearly to superpositions of positive and negative
frequency solutions. (Alternatively, we may declare positive and
negative frequency solutions, respectively, to define two
superselection sectors. The procedure here is analogous to
\cite{GenRepIn}.)  This completes the construction of the physical
Hilbert space.

From physical states, $\beta_+$-dependent expectation values (or
moments) can be computed, playing the role of evolving observables.
As with the non-relativistic harmonic oscillator, this is most easily
done using ladder operators: $\hat{\alpha} =
\sqrt{\hbar/2} \left( \hat{a}^{\dagger} + \hat{a} \right)$,
$\hat{p}_{\alpha} = i\sqrt{\hbar/2} \left( \hat{a}^{\dagger} -
\hat{a} \right)$. A direct calculation gives
\begin{eqnarray}
\langle \Psi_+, \hat{\alpha} \Psi_+ \rangle (\beta_+) &=& \sum_{n=0}^{\infty}
\sqrt{2\hbar(n+1)} {\rm Re} \left( \bar{c}_n
c_{n+1} \exp \left( -i\beta_+(\lambda_{n+1} - \lambda_{n})/\hbar
\right) \right) \label{alphaExp}\\
\langle \Psi_+, \hat{p}_{\alpha} \Psi_+ \rangle (\beta_+) &=&
\sum_{n=0}^{\infty} \sqrt{2\hbar(n+1)} {\rm Im}
\left( \bar{c}_n c_{n+1} \exp \left( -i\beta_+(\lambda_{n+1} -
\lambda_{n})/\hbar \right) \right) \label{pExp}
\end{eqnarray}
and similarly for moments.

Physical Hilbert spaces can be constructed and physical states
decomposed in this way whenever one knows an explicit diagonalization
of the Hamiltonian $\hat{H}$. For our model, the closeness to the
harmonic oscillator has an additional advantage in that it allows us
to use its simple form of coherent states --- Gaussians of arbitrary
width expanded in the stationary states --- as initial values for
evolution in $\beta_+$. For the non-relativistic harmonic oscillator,
the resulting physical states would be dynamical coherent states:
their shape would remain unchanged and they keep saturating the
uncertainty relation at all times. Moreover, their expectation values
follow the classical trajectories exactly, without quantum
back-reaction.

For the relativistic harmonic oscillator, and thus our anisotropic toy
model, the dynamical behavior of the states is still to be seen. We
thus assume an initial state, at some fixed value of $\beta_+$, of the
kinematical coherent form:
\begin{equation}\label{cn}
c_n = \exp \left( -\frac{|z|^2}{2} \right)
\frac{z^n}{\sqrt{n!}}\quad, \quad z \in {\mathbb C}
\end{equation}
such that $\Psi(\alpha,0)=
(2/\pi)^{1/4}\exp\left(-\frac{1}{2}(|z|^2-z^2+2\alpha^2-4iz\alpha)\right)$
from (\ref{States}).
A time-dependent physical state then has coefficients
\[
 c_n e^{-i\lambda_n \beta_+/\hbar}=\frac{1}{\sqrt{n!}}e^{-|z|^2/2}z^n
e^{-i\sqrt{2n+1}\beta_+/\sqrt{\hbar}}
\]
in its expansion by the $\varphi_n(\alpha)$. With the square root of
$2n+1$, the exponentials cannot simply be combined to a
$\beta_+$-dependent $z(\beta_+)^n$. The shape of the state changes as
$\beta_+$ moves away from zero: the time-dependent coefficients are no
longer of the form (\ref{cn}) for $\beta_+\not=0$. Physical
states with initial conditions given by the coherent states of the
non-relativistic harmonic oscillator are not dynamical coherent
states. The model introduced here does show spreading and quantum
back-reaction, which makes it interesting for a comparison with
effective constraint techniques.

\subsection{Effective constraints}
\label{sec:eff}

In an effective treatment of a quantum system, we can consider the
same dynamics, but focus on the algebra of observables. Results
will thus be manifestly representation independent. To set up this
framework, no approximations are required; we are thus dealing with an
exact quantum theory. Only when evaluating the equations, which would
give us expressions such as $\langle\hat{\alpha}\rangle(\beta_+)$ or
moments of a state as functions of $\beta_+$, do approximation schemes
typically enter. This is no difference to a representation dependent
treatment, where exact evaluations of expectation values such as
(\ref{alphaExp}) or (\ref{pExp}) are hard to sum explicitly.

At the kinematical level, effective techniques are based entirely on
the algebra of basic operators, in our case
$[\hat{\beta}_+,\hat{p}_+]=i\hbar$ and
$[\hat{\alpha},\hat{p}_{\alpha}]=i\hbar$, with all other basic
commutators vanishing. As it happens at the quantum level, dynamics is
brought in by a constraint operator $\hat{C}$ which might have more
complicated algebraic relationships with the basic operators, no
longer forming a closed algebra.

A whole representation of these algebraic relationships on wave
functions carries much more details than necessary for extracting
physical results. Instead, it is often convenient to focus directly on
expectation values and derive dynamical equations for them, avoiding
the detour of computing wave functions. Expectation values are not
sufficient to characterize a state or its dynamics, but when combined
with all moments
\begin{equation}
 \Delta(O_1 \ldots O_n):= \left\langle
 \prod_{i=1}^n(\hat{O}_i-\langle\hat{O}_i\rangle)\right\rangle_{\rm
 Weyl}
\end{equation}
a complete set of variables results. Here, the subscript ``Weyl'' in
the definition of the moments indicates that we are ordering all
operator products totally symmetrically. Any of the basic operators,
$\hat{\beta}_+$, $\hat{p}_+$, $\hat{\alpha}$ and $\hat{p}_{\alpha}$,
can appear as the $\hat{O}_i$ in the basic moments. (To match with
standard notation, we will write $\Delta(O^2)= (\Delta O)^2$ for
fluctuations.)

Expectation values of basic operators together with the moments can be
used to characterize an arbitrary state (pure or mixed); they can be
used as coordinates on the state space. Moreover, the commutator of
basic operators endows the state space with a Poisson structure,
defined by
\begin{equation}
 \{\langle\hat{A}\rangle,\langle\hat{B}\rangle\}:=
 \frac{\langle[\hat{A},\hat{B}]\rangle}{i\hbar}
\end{equation}
for any operators $\hat{A}$ and $\hat{B}$. By linearity and the
Leibniz rule, this defines Poisson brackets between all the moments
and expectation values. In this way, the quantum phase space is
defined. It is not a linear space since there are restrictions for the
moments, most importantly the uncertainty relations such as
\begin{equation}
 (\Delta\alpha)^2(\Delta p_{\alpha})^2- (\Delta(\alpha
 p_{\alpha})^2\geq \frac{\hbar^2}{4}\,.
\end{equation}

On this kinematical quantum phase space, the constraint $\hat{C}$ must
be imposed. For a constraint operator polynomial in the basic
operators,
\begin{equation}
 C_{\rm pol}:= \langle (\widehat{\rm pol}-\langle\widehat{\rm
   pol}\rangle) \hat{C}\rangle
\end{equation}
defines a function on the state space, expandable in the moments, for
any arbitrary polynomial $\widehat{\rm pol}$ of the basic
operators. This infinite set of functions\footnote{A similar setup can
be used in BRST quantizations, where the BRST charge would replace the
constraint operators \cite{HamBRST}.}  satisfies two important
properties: (i) all these functions vanish on physical states
annihilated by $\hat{C}$, and (ii) they form a first class set in the
sense of classical constraint analysis, i.e.\ $\{C_{\rm pol},C_{{\rm
pol}'}\}$ vanishes on the subset of the phase space where all $C_{\rm
pol}$ vanish. The quantum constraint $\hat{C}$ can thus be implemented
on the space of moments by imposing the infinite set $\{C_{\rm pol}\}$
as constraints as one would do it on a classical phase space. We have
to find the submanifold on which all constraints vanish, and factor
out the flow generated by them. If this reduction is completed, we
obtain the physical quantum phase space and can look for solutions of
observables.

This procedure has several advantages \cite{EffCons}. As already
mentioned, it is completely representation independent and instead
focuses on algebraic aspects of a quantum system. As a consequence,
implementing constraint operators with zero in their continuous
spectra is no different from implementing those with zero in their
discrete spectrum. Any difficulties in finding physical inner products
can be avoided, for the physical normalization arises automatically
when the constraints are solved for moments.

Once we try to find specific solutions, there are of course practical
difficulties. We are dealing with infinitely many constraints on an
infinite-dimensional phase space. Sometimes, this set of equations
decouples to finitely many ones, but this happens only in rare
solvable cases. In more general systems, we must use approximations to
reduce the set to a finite one of relevant equations, with a
semiclassical approximation as the main example. Here, we look for
solutions whose moments satisfy a certain hierarchy, higher moments in
the semiclassical case being suppressed by powers of $\hbar$,
$\Delta(O_1\ldots O_n)\propto \hbar^{n/2}$. To any given order in
$\hbar$, only a finite number of moments need be considered, subject
to a finite number of non-trivial constraints.

In our model, to second order in moments we have the effective
constraints
\begin{eqnarray}
C&=& \langle\hat{p}_+\rangle^2 - \langle\hat{p}_{\alpha}\rangle^2
- \langle\hat{\alpha}\rangle^2 + (\Delta p_+)^2 - (\Delta p_{\alpha})^2 -
(\Delta \alpha)^2 = 0 \\
C_{\beta_+}& =& 2\langle\hat{p}_+\rangle \Delta(\beta_+ p_+) + i\hbar
\langle\hat{p}_+\rangle - 2\langle\hat{p}_{\alpha}\rangle
\Delta(\beta_+ p_{\alpha}) - 2\langle\hat{\alpha}\rangle
\Delta(\beta_+\alpha) = 0 \\
C_{p_+}& =&
2\langle\hat{p}_+\rangle (\Delta p_+)^2 -
2\langle\hat{p}_{\alpha}\rangle  \Delta(p_+p_{\alpha})
-2\langle\hat{\alpha}\rangle\Delta(p_+\alpha) = 0\\
C_{\alpha} & =& 2\langle\hat{p}_+\rangle \Delta(p_+\alpha) -
2\langle\hat{p}_{\alpha}\rangle \Delta(\alpha p_{\alpha})
- i\hbar \langle\hat{p}_{\alpha}\rangle-
2\langle\hat{\alpha}\rangle(\Delta \alpha)^2 = 0 \\
C_{p_{\alpha}}& =& 2\langle\hat{p}_+\rangle \Delta(p_+p_{\alpha}) -
2\langle\hat{p}_{\alpha}\rangle(\Delta p_{\alpha})^2
-2\langle\hat{\alpha}\rangle \Delta(\alpha p_{\alpha}) +
i\hbar \langle\hat{\alpha}\rangle = 0\,. \label{eq:2_ord_constr_pot}
\end{eqnarray}
The reduction has been performed in \cite{EffConsRel}, with the result
that the physical state space is equivalent to a deparametrized
quantum system with constraint $C_Q=\langle\hat{p}_+\rangle\pm H_Q$
with the reduced Hamiltonian
\begin{eqnarray}
H_Q &=& \sqrt{\langle\hat{p}_{\alpha}\rangle^2 + \langle\hat{\alpha}\rangle^2}
\Biggl( 1 + \frac{ \langle\hat{\alpha}\rangle^2(\Delta p_{\alpha} )^2 -
2\langle\hat{\alpha}\rangle \langle\hat{p}_{\alpha}\rangle
\Delta(\alpha p_{\alpha}) +
\langle\hat{p}_{\alpha}\rangle^2
(\Delta \alpha)^2}{2(\langle\hat{p}_{\alpha}\rangle^2
+ \langle\hat{\alpha}\rangle^2)^2} \Biggr) \\
&&+ O\left((\Delta p_{\alpha})^4\right) +
O\left((\Delta \alpha)^4\right) +
O\left((\Delta (\alpha p_{\alpha}))^2\right) \,.
\label{eq:E_space_pot}
\end{eqnarray}

To leading order, we reproduce the classical Hamiltonian, but there
are corrections from quantum fluctuations and correlations coupling to
the expectation values. Moreover, from the Poisson brackets between
moments we obtain Hamiltonian equations of motion telling us how a
state spreads or is being squeezed. We have
\begin{eqnarray}
\frac{\rm d \langle\hat{\alpha}\rangle }{{\rm d} \beta_+} &=&
\frac{\langle\hat{p}_{\alpha}\rangle}{\sqrt{\langle\hat{p}_{\alpha}\rangle^2
+ \langle\hat{\alpha}\rangle^2}}\label{eq:eff1st}\\  && +
\frac{\langle\hat{p}_{\alpha}\rangle (\Delta \alpha)^2 \left(
2\langle\hat{\alpha}\rangle^2 -
  \langle\hat{p}_{\alpha}\rangle^2\right) + \langle\hat{\alpha}\rangle
 \Delta(\alpha p_{\alpha})
\left( 4\langle\hat{p}_{\alpha}\rangle^2 -
2\langle\hat{\alpha}\rangle^2\right) -
3\langle\hat{p}_{\alpha}\rangle\langle\hat{\alpha}\rangle^2
(\Delta p_{\alpha})^2}{2 \left( \langle\hat{p}_{\alpha}\rangle^2 +
\langle\hat{\alpha}\rangle^2 \right)^{\frac{5}{2}}}\nonumber \\
\frac{\md \langle\hat{p}_{\alpha}\rangle }{{\rm d} \beta_+} &=&
\frac{-\langle\hat{\alpha}\rangle}{\sqrt{\langle\hat{p}_{\alpha}\rangle^2
+ \langle\hat{\alpha}\rangle^2}}\\
&& +
\frac{3\langle\hat{\alpha}\rangle \langle\hat{p}_{\alpha}\rangle^2
(\Delta \alpha)^2  - \langle\hat{p}_{\alpha}\rangle
  \Delta(\alpha p_{\alpha}) \left(
4\langle\hat{\alpha}\rangle^2 -
2\langle\hat{p}_{\alpha}\rangle^2\right)
- \langle\hat{\alpha}\rangle (\Delta p_{\alpha})^2
\left( 2\langle\hat{p}_{\alpha}\rangle^2 -
\langle\hat{\alpha}\rangle^2\right)}{2 \left(
\langle\hat{p}_{\alpha}\rangle^2 +
\langle\hat{\alpha}\rangle^2\right)^{\frac{5}{2}}} \nonumber \\
\frac{\md (\Delta\alpha)^2}{{\rm d} \beta_+} &=&
\frac{2\langle\hat{\alpha}\rangle^2\Delta(\alpha p_{\alpha}) -
2\langle\hat{\alpha}\rangle \langle\hat{p}_{\alpha}\rangle
(\Delta \alpha)^2}{\left( \langle\hat{p}_{\alpha}\rangle^2 +
\langle\hat{\alpha}\rangle^2\right)^{\frac{3}{2}}}\\
\frac{\md (\Delta p_{\alpha})^2}{{\rm d} \beta_+} &=&
\frac{2 \langle\hat{\alpha}\rangle \langle\hat{p}_{\alpha}\rangle
(\Delta p_{\alpha})^2 -
  2\langle\hat{\alpha}\rangle^2\Delta(\alpha p_{\alpha})}{\left(
\langle\hat{p}_{\alpha}\rangle^2 +
\langle\hat{\alpha}\rangle^2\right)^{\frac{3}{2}}}
\\ \label{eq:efflast} \frac{\md \Delta(\alpha p_{\alpha}) }{{\rm d}
\beta_+} &=& \frac{ \langle\hat{\alpha}\rangle^2(\Delta
p_{\alpha})^2 - \langle\hat{p}_{\alpha}\rangle^2(\Delta
\alpha)^2}{\left( \langle\hat{p}_{\alpha}\rangle^2 +
\langle\hat{\alpha}\rangle^2 \right)^{\frac{3}{2}}}\,.
\end{eqnarray}
Solutions $\langle\hat{\alpha}\rangle(\beta_+)$,
$\langle\hat{p}_\alpha\rangle(\beta_+)$, $(\Delta\alpha)^2(\beta_+)$,
$(\Delta p_{\alpha})^2(\beta_+)$ and $\Delta(\alpha
p_{\alpha})(\beta_+)$ provide physical observables, corresponding to
expectation values and moments in physical states, telling us how a
state moves and spreads.

\section{Comparison}

Initially, the kinematical coherent state with expansion
coefficients given by~(\ref{cn}) yields the expectation values
$\langle \hat{\alpha} \rangle|_{t=0} = {\rm Re}(z)$\ and $ \langle
\hat{p}_{\alpha} \rangle|_{t=0} = {\rm Im}(z)$. The second order
moments then saturate the uncertainty relation and their values
are
\[
(\Delta \alpha)^2|_{t=0} = \frac{\hbar}{2}, \quad \quad (\Delta
p_{\alpha})^2|_{t=0} = \frac{\hbar}{2}, \quad \quad \Delta (\alpha
p_{\alpha})|_{t=0} = 0.
\]

\begin{center}
\begin{figure}[htbp!]
\includegraphics[width=12cm]{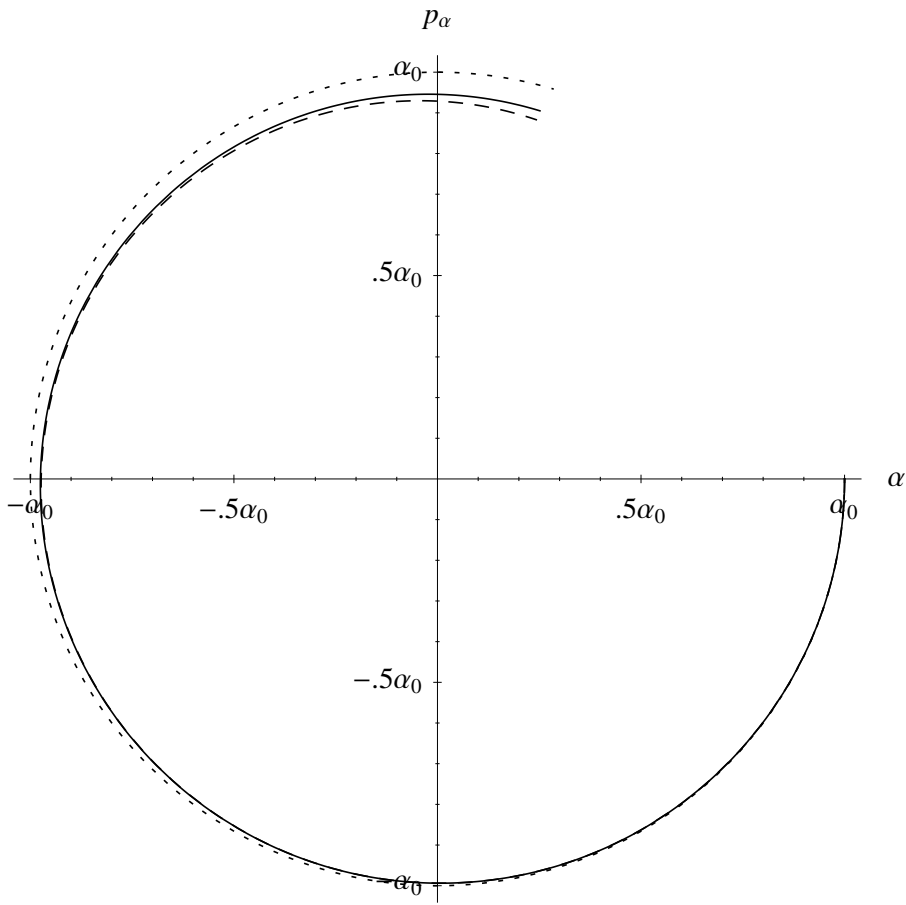}
\caption{\label{fig:1} Classical (dotted), coherent state (solid)
and effective (dashed) phase space trajectories, evolved for $0 \leq
\beta_+ \leq 5\alpha_0$.}
\end{figure}
\end{center}

For a concrete comparison we select a state that is initially peaked
about $\langle\hat{\alpha}\rangle = \alpha_0$\ and
$\langle\hat{p}_{\alpha}\rangle = 0$, so that $z = \alpha_0$. In order
for the state to be semiclassical, we need $\alpha_0$\ to be
``significantly larger'' than $\sqrt{\hbar}$. We make a concrete
choice $\alpha_0 = 10\hbar^{\frac{1}{2}}$. In Fig.~\ref{fig:1} we
plot alongside each other the classical and two quantum-corrected
trajectories of the system starting from the above initial values. The
two corrected trajectories were calculated using two different
methods: the kinematical coherent state of Section~\ref{sec:state} and
effective equations truncated at order $\hbar$,
Eqs.~(\ref{eq:eff1st})--(\ref{eq:efflast}) of Section~\ref{sec:eff}.

\begin{center}
\begin{figure}[htbp!]
\includegraphics[width=12cm]{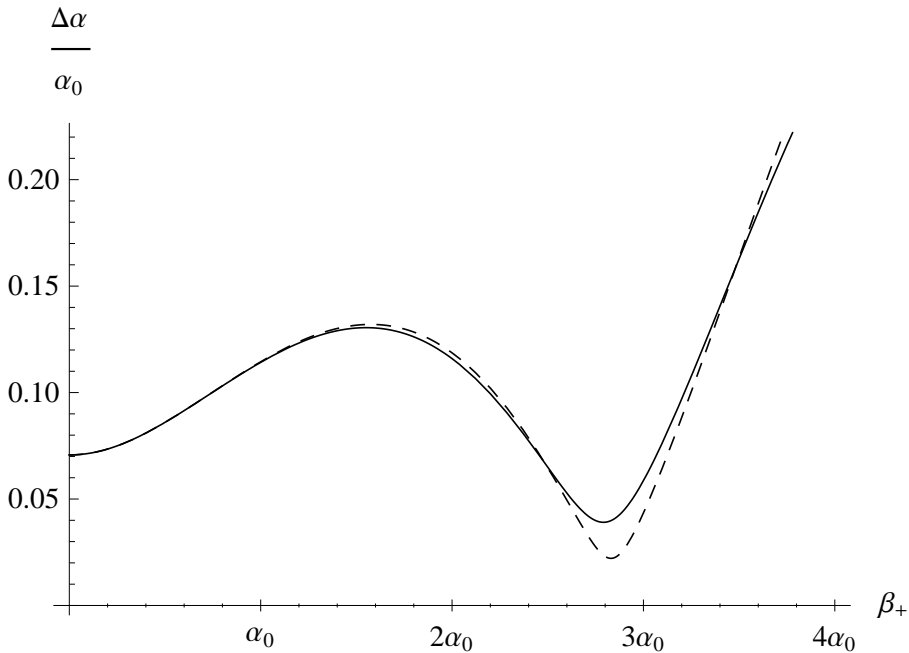}
\caption{\label{fig:2a} Coherent state (solid) and effective
(dashed) evolution of the second order moment $\Delta \alpha =
\sqrt{(\Delta \alpha)^2}$\ in units of $\alpha_0$.}
\end{figure}
\end{center}

\begin{center}
\begin{figure}[htbp!]
\includegraphics[width=12cm]{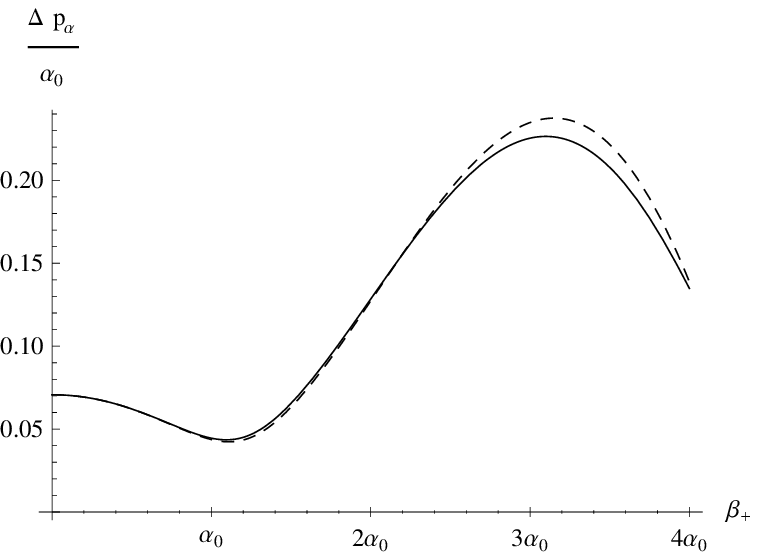}
\caption{\label{fig:2b} Coherent state (solid) and effective
(dashed) evolution of the second order moment $\Delta p_{\alpha} =
\sqrt{(\Delta p_{\alpha})^2}$\ in units of $q_0$.}
\end{figure}
\end{center}

\begin{center}
\begin{figure}[htbp!]
\includegraphics[width=12cm]{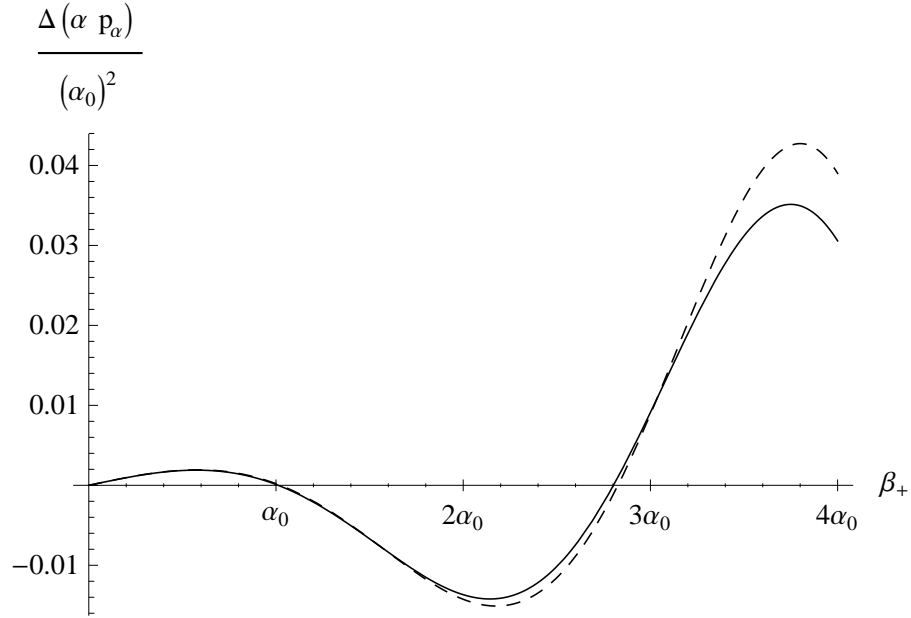}
\caption{\label{fig:2c} Coherent state (solid) and effective
(dashed) evolution of the second order moment $\Delta (\alpha
p_{\alpha})$\ in units of $\alpha_0^2$.}
\end{figure}
\end{center}

The two quantum trajectories agree very well for much of the
evolution shown. In Figs.~\ref{fig:2a}--\ref{fig:2c} we plot the
quantum evolution of the second order moments generated by
$\hat{\alpha}$\ and $\hat{p}_{\alpha}$. From Fig.~\ref{fig:2b} in
particular it is clear that the semiclassical approximation breaks
down somewhere between $\beta_+=2\alpha_0$\ and $\beta_+=3\alpha_0$\
as $\Delta p_{\alpha}$\ is no longer ``much smaller'' than
$\alpha_0$. Up until that point both methods for quantum
evolution are in close agreement in describing not only the
trajectory in the $\alpha-p_{\alpha}$\ space but also the evolution
of the second order moments themselves. Since semiclassicality was
used to obtain the truncated system of equations, there is no reason
to expect it to be accurate beyond $\beta_+=3\alpha_0$.

The two methods agree excellently where the domains of their
applicability overlap. Both methods have their own strengths and
weaknesses. The weakness of the semiclassically truncated effective
equations should by now stand out --- in some systems semiclassicality
eventually breaks down and moments of high order dominate. Direct
evaluation on the states does not rely on this approximation and
remains a valid method for computing the wave function at all
times. The shortcomings of the latter technique are less obvious and
are of practical nature. In order to apply the method one needs, first
of all, to decompose the initial state as a sum of the eigenstates of
the Hamiltonian, which for an arbitrary state and a typical
Hamiltonian can be complicated. During evolution, each of the
eigenstates acquires a phase factor and they need to be re-summed to
compute the wave function at a later time.  For an arbitrary state, the
sum may converge very slowly requiring one to sum over a very large
number of eigenstates to obtain an accurate description of the
wave function. Finally, for expectation values and moments of
observables further integrations must be done. In systems of several
degrees of freedom, this will add considerably to computation times.

If one is to make robust predictions, a range of initially
semiclassical states with a variety of initial values of moments
should be considered --- the procedure is very complicated to
implement using state decomposition but amounts to nothing more than
simply changing the initial conditions in the case of the effective
equations. In the subsections that follow we use the methods
separately and exploit their individual strengths.

\subsection{Long-term behavior of the state}

\begin{center}
\begin{figure}[htbp!]
\includegraphics[width=12cm]{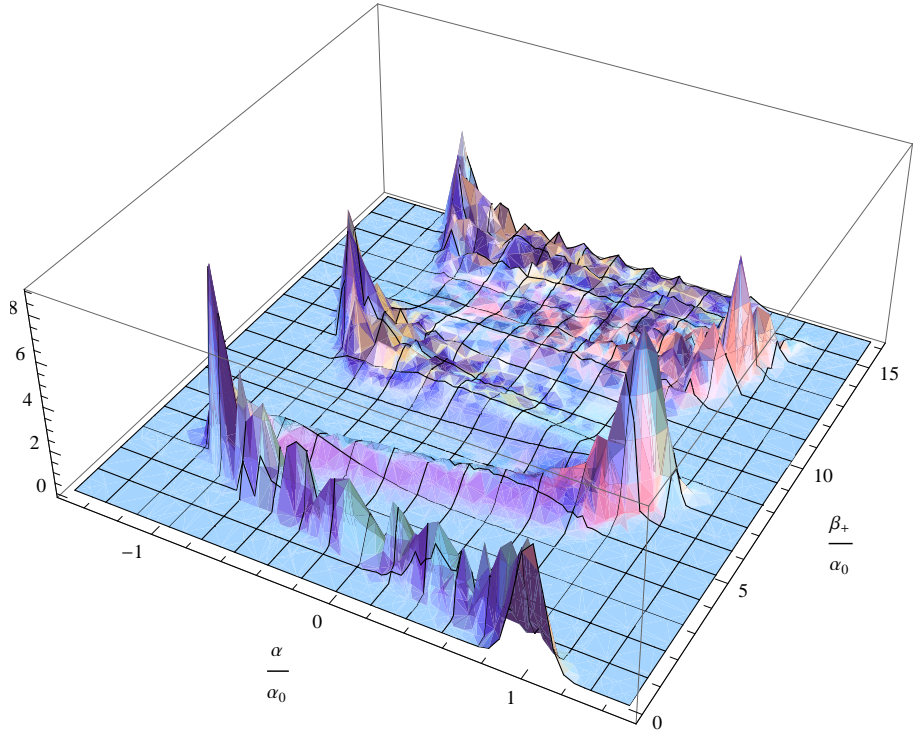}
\caption{\label{fig:3} Square-magnitude of the coherent state
wave function evolved for $0 < \beta_{+} < 15\alpha_0$.}
\end{figure}
\end{center}

Knowing the state exactly, allows us to plot the magnitude of the
wave function and make precise long-term predictions. From
Fig.~\ref{fig:3} we see that after $\beta_+\approx 10\alpha_0$\ the
state becomes highly quantum, spread out over an entire
orbit. Expectation values can no longer be interpreted as the most
probable outcome of a measurement.

\begin{center}
\begin{figure}[htbp!]
\includegraphics[width=12cm]{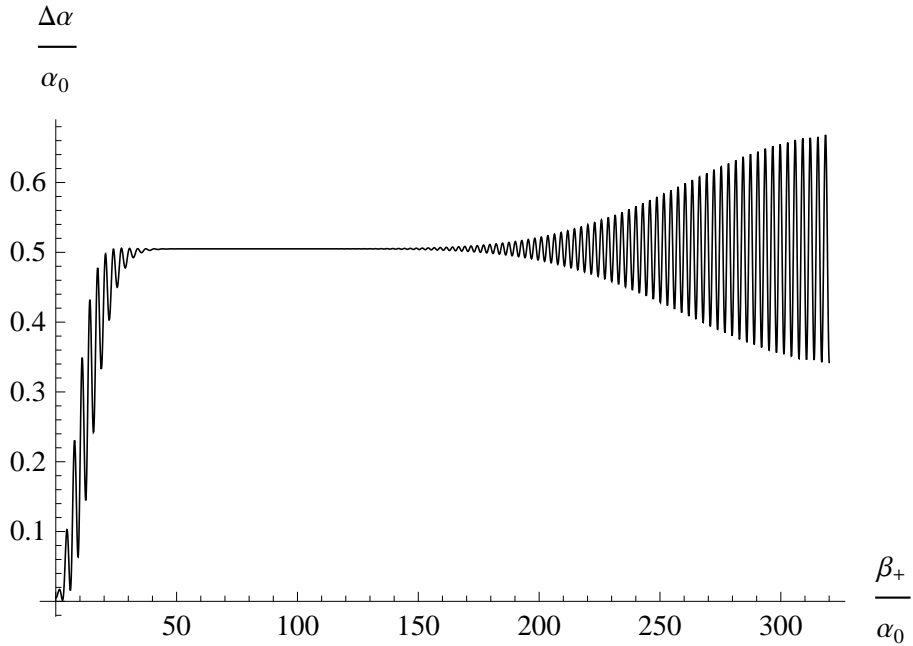}
\caption{\label{fig:4a} Coherent state evolution of the second order
moment $\Delta \alpha = \sqrt{(\Delta \alpha)^2}$\ in units of
$\alpha_0$.}
\end{figure}
\end{center}

\begin{center}
\begin{figure}[htbp!]
\includegraphics[width=12cm]{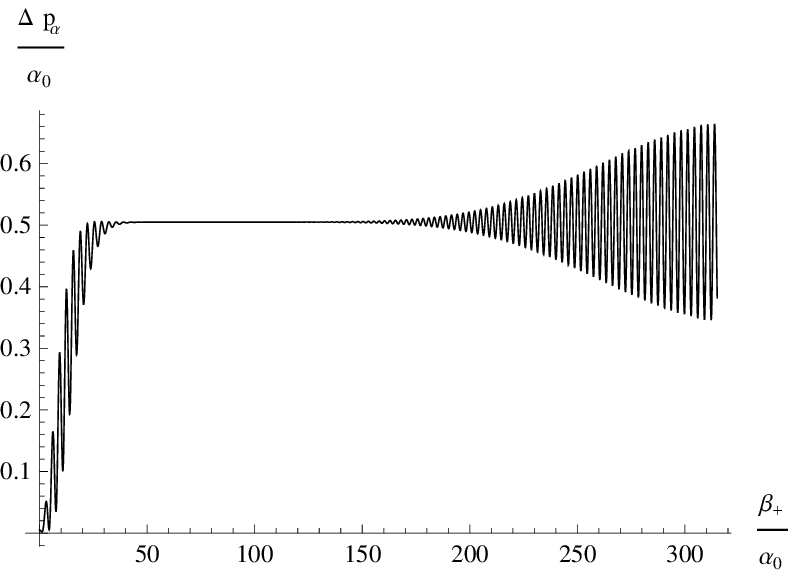}
\caption{\label{fig:4b} Coherent state (solid) and effective
(dashed) evolution of the second order moment $\Delta p_{\alpha} =
\sqrt{(\Delta p_{\alpha})^2}$\ in units of $\alpha_0$.}
\end{figure}
\end{center}

\begin{center}
\begin{figure}[htbp!]
\includegraphics[width=12cm]{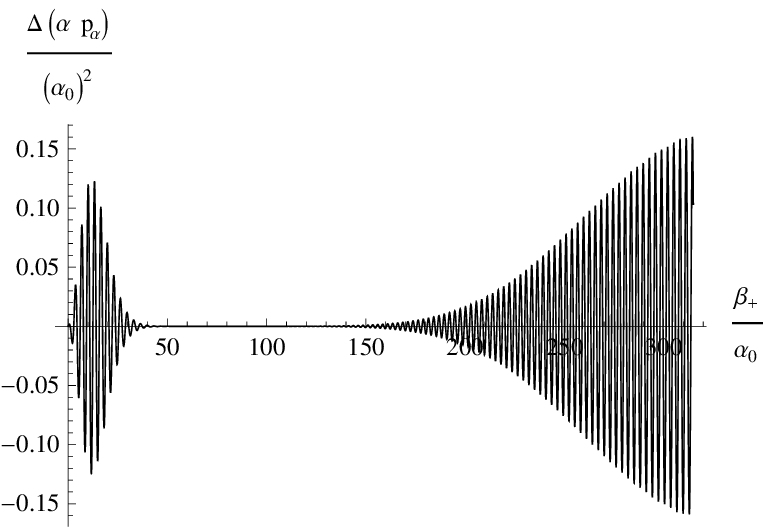}
\caption{\label{fig:4c} Coherent state (solid) and effective
(dashed) evolution of the second order moment $\Delta (\alpha
p_{\alpha})$\ in units of $\alpha_0$.}
\end{figure}
\end{center}

\begin{center}
\begin{figure}[htbp!]
\includegraphics[width=12cm]{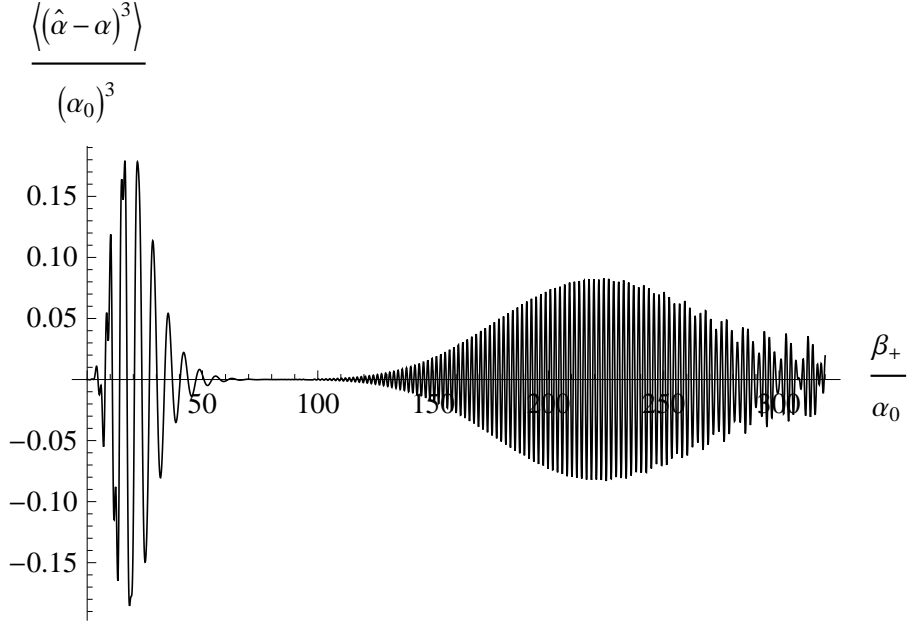}
\caption{\label{fig:5} Coherent state evolution of the third order
moment $\langle (\hat{\alpha} - \langle\hat{\alpha}\rangle)^3
\rangle$\ in units of $\alpha_0$.}
\end{figure}
\end{center}

In Figs.~\ref{fig:4a}--\ref{fig:4c} we look at the long term behavior
of the leading order quantum degrees of freedom. The magnitudes of
$(\Delta \alpha)^2$\ and $(\Delta p_{\alpha})^2$\ rise rapidly until
around $\beta_+=40\alpha_0$. The state remains stable during
$40\alpha_0 < \beta_+ < 200 \alpha_0$\ after which all second order
moments begin to oscillate with an increasing amplitude. The amplitude
reaches its peak around $\beta_+=300\alpha_0$; thereafter the moments
keep oscillating with a stable amplitude for as long as the evolution
has been traced, up to around $\beta_+ = 10^5 \alpha_0$. As can be
seen from the example of $\langle (\hat{\alpha} -
\langle\hat{\alpha}\rangle)^3 \rangle$\ in Fig.~\ref{fig:5}, third
order moments follow a similar pattern.

\begin{center}
\begin{figure}[htbp!]
\includegraphics[width=12cm]{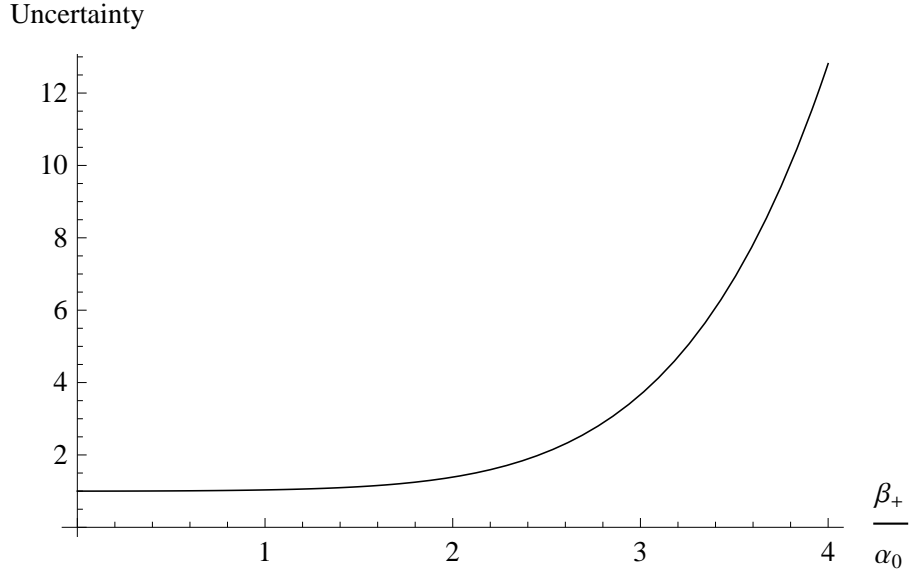}
\caption{\label{fig:6} Short-term coherent state evolution of the
``quantum uncertainty'' defined as $(\Delta \alpha)^2(\Delta
p_{\alpha})^2 - \left( \Delta (\alpha p_{\alpha})\right)^2$\ in
units of $\frac{\hbar^2}{4}$.}
\end{figure}
\end{center}

\begin{center}
\begin{figure}[htbp!]
\includegraphics[width=12cm]{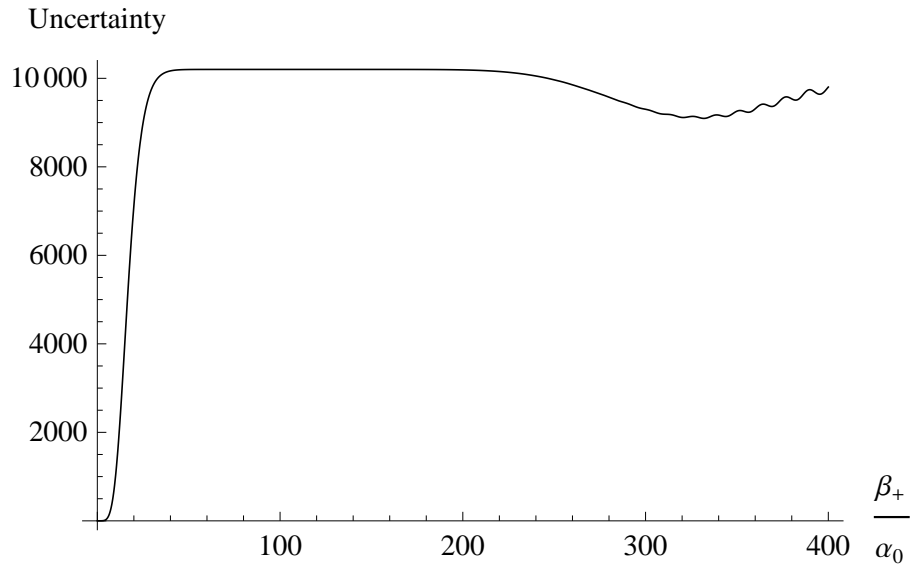}
\caption{\label{fig:6b} Long-term coherent state evolution of the
``quantum uncertainty'' defined as $(\Delta \alpha)^2(\Delta
p_{\alpha})^2 - \left( \Delta (\alpha p_{\alpha})\right)^2$\ in
units of $\frac{\hbar^2}{4}$.}
\end{figure}
\end{center}

As expected, semiclassicality eventually breaks down and quantum
fluctuations become large; this trend is illustrated by the evolving
uncertainty measure $(\Delta \alpha)^2(\Delta p_{\alpha})^2 - \left(
\Delta (\alpha p_{\alpha})\right)^2$ (bounded below by the uncertainty
relation) shown in Fig.~\ref{fig:6}. After the initial increase, there
is a period of approximate stability; then the leading order
fluctuations start to oscillate with large amplitudes. Even though
some moments return to small values during these oscillations,
semiclassicality is not regained as shown be the long-term behavior of
the uncertainties in Fig.~\ref{fig:6b}. Details of this behavior,
found numerically, appears rather characteristic, but it is difficult
to find an explanation based on the underlying equations.

\subsection{Short-term evolution with varying initial conditions}

\begin{center}
\begin{figure}[htbp!]
\includegraphics[width=12cm]{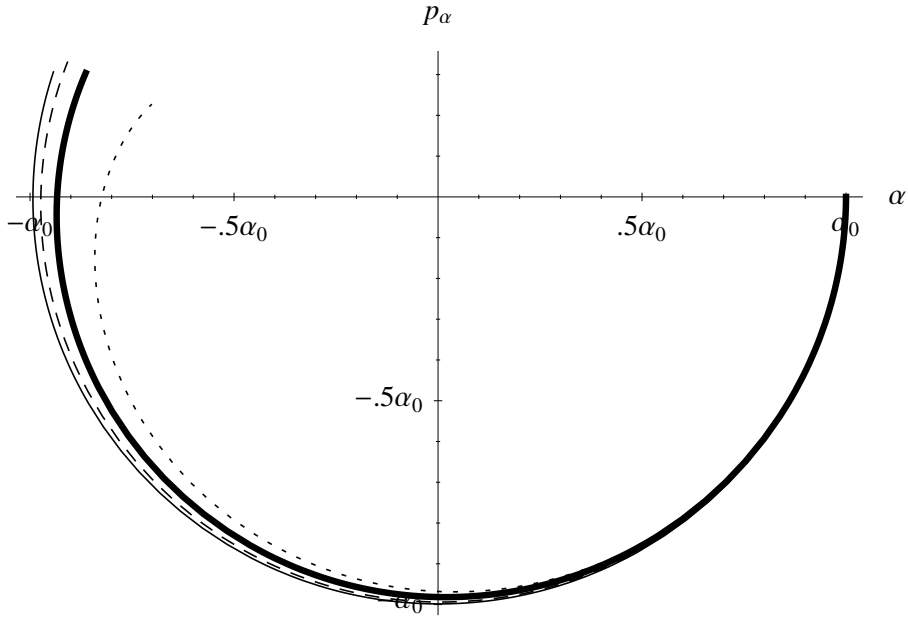}
\caption{\label{fig:7} Effective evolution of semiclassical states
with different initial values of moments for $0 < \beta_+ <
3.5\alpha_0$. The initial values of moments are as follows. Dashed
line---the original coherent state: $(\Delta \alpha)^2=(\Delta
p_{\alpha})^2 = .005 \alpha_0^2$, $\Delta (\alpha p_{\alpha}) = 0$.
Dotted line: $(\Delta \alpha)^2=.025\alpha_0^2$, $(\Delta
p_{\alpha})^2 = .001 \alpha_0^2$, $\Delta (\alpha p_{\alpha}) = 0$.
Thin solid line: $(\Delta \alpha)^2=.001\alpha_0^2$, $(\Delta
p_{\alpha})^2 = .025 \alpha_0^2$, $\Delta (\alpha p_{\alpha}) = 0$.
Thick solid line:$(\Delta \alpha)^2=.008 \alpha_0^2$, $(\Delta
p_{\alpha})^2 = .008 \alpha_0^2$, $\Delta (\alpha p_{\alpha}) = .005
\alpha_0^2$.}
\end{figure}
\end{center}

\begin{center}
\begin{figure}[htbp!]
\includegraphics[width=12cm]{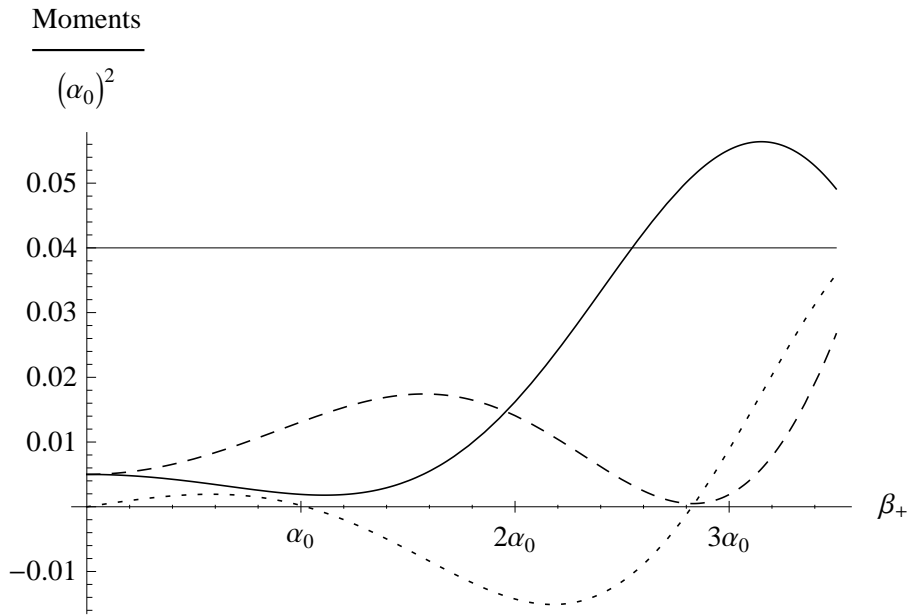}
\caption{\label{fig:8a} Effective evolution of second order moments
$(\Delta \alpha)^2$\ (dashed), $(\Delta p_{\alpha})^2$\ (solid) and
$\Delta (\alpha p_{\alpha})$\ (dotted), which initially take the
values $.005 \alpha_0^2$, $.005\alpha_0^2$\ and $0$\ respectively.}
\end{figure}
\end{center}

\begin{center}
\begin{figure}[htbp!]
\includegraphics[width=12cm]{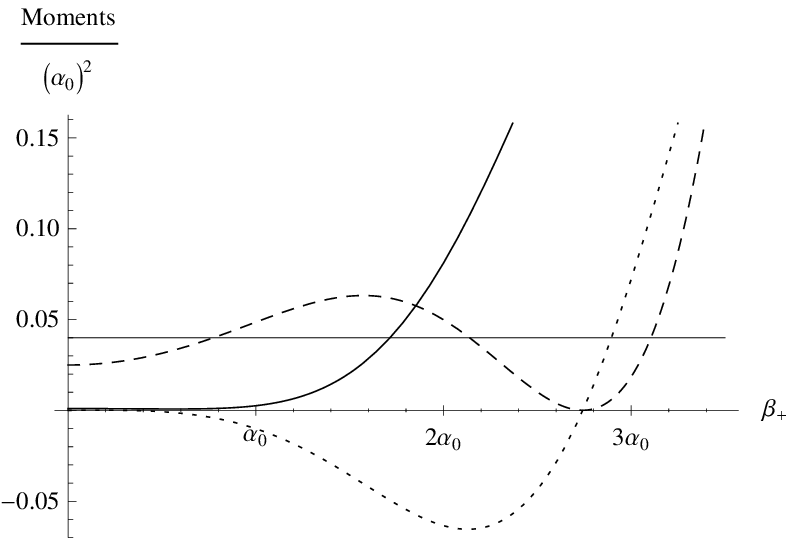}
\caption{\label{fig:8b} Effective evolution of second order moments
$(\Delta \alpha)^2$\ (dashed), $(\Delta p_{\alpha})^2$\ (solid) and
$\Delta (\alpha p_{\alpha})$\ (dotted), which initially take the
values $.025 \alpha_0^2$, $.001\alpha_0^2$\ and $0$\ respectively.}
\end{figure}
\end{center}

\begin{center}
\begin{figure}[htbp!]
\includegraphics[width=12cm]{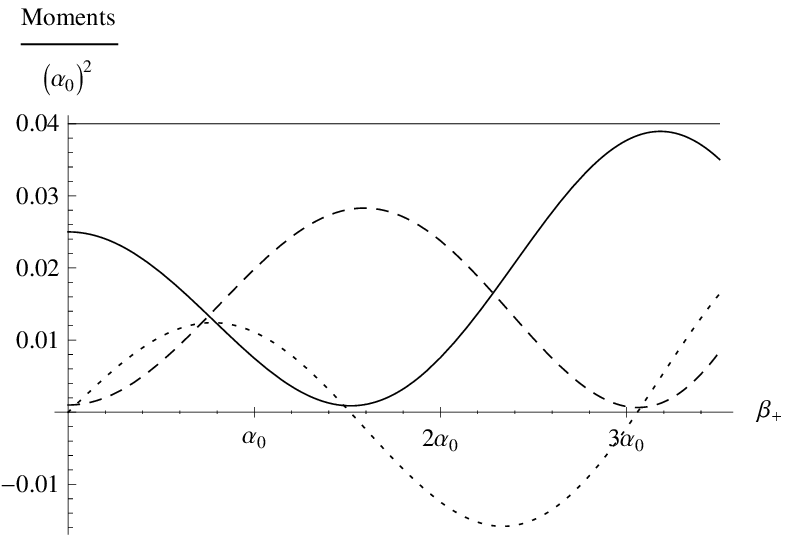}
\caption{\label{fig:8c} Effective evolution of second order moments
$(\Delta \alpha)^2$\ (dashed), $(\Delta p_{\alpha})^2$\ (solid) and
$\Delta (\alpha p_{\alpha})$\ (dotted), which initially take the
values $.001 \alpha_0^2$, $.025\alpha_0^2$\ and $0$\ respectively.}
\end{figure}
\end{center}

\begin{center}
\begin{figure}[htbp!]
\includegraphics[width=12cm]{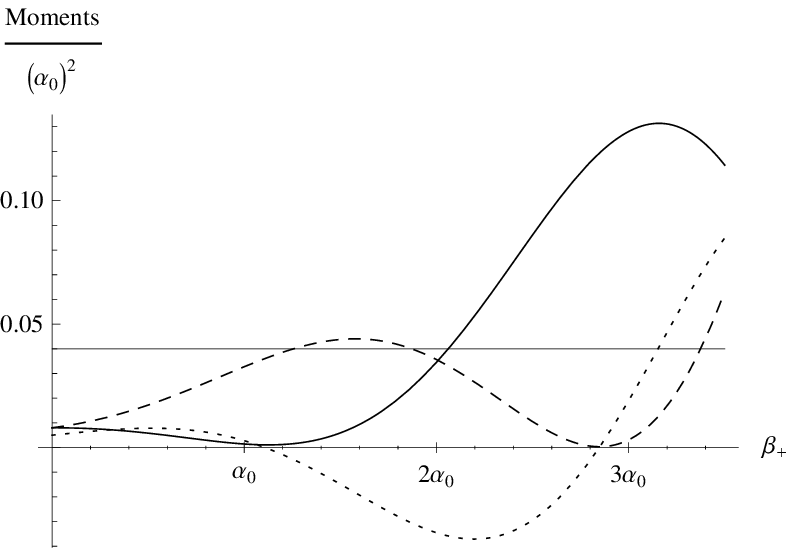}
\caption{\label{fig:8d} Effective evolution of second order moments
$(\Delta \alpha)^2$\ (dashed), $(\Delta p_{\alpha})^2$\ (solid) and
$\Delta (\alpha p_{\alpha})$\ (dotted), which initially take the
values $.008 \alpha_0^2$, $.008\alpha_0^2$\ and $.005\alpha_0^2$\
respectively.}
\end{figure}
\end{center}

We now evolve effectively starting from the same initial expectation
values, but varying the initial values of second order moments.  By
the specific choices, some sets of moments used here no longer
saturate the uncertainty relations, and some have non-vanishing
correlations. They cannot correspond to Gaussian states, and so their
initial configurations would be much more difficult to implement using
wave functions of physical states.  The results are plotted in
Fig.~\ref{fig:7}. In Figs.~\ref{fig:8a}--\ref{fig:8d} we plot the
corresponding effective evolution of moments for each set of the
initial conditions, where the thin horizontal line approximately
indicates the threshold of the semiclassical approximation for the
second order moments at $.04\alpha_0^2$.

Larger $(\Delta \alpha)^2$\ results in a larger deviation from the
classical behavior and a faster breakdown of the semiclassical
approximation. The above effect is much less sensitive to the momentum
spread $(\Delta p_{\alpha})^2$. This disparity between the effects of
the two spreads may seem surprising given the symmetry between
$\hat{\alpha}$\ and $\hat{p}_{\alpha}$\ in the expression for the
physical Hamiltonian. This symmetry, however is broken by the initial
state we have chosen, which is peaked about
$\langle\hat{p}_{\alpha}\rangle = 0$\ and $\langle\hat{\alpha}\rangle
= \alpha_0$, so that the spread in $\alpha$, produces a larger spread
in the energy than the spread in $p_{\alpha}$.

\section{Effective evolution of a radiation-filled universe}
\label{sec:rad_universe}

In this section we use the effective equations on their own to analyze
quantum corrections to the dynamics of a radiation-filled Bianchi I
universe briefly mentioned in Section~\ref{sec:model}.  This model has
a more realistic matter content than the one analyzed above. We recall
that the energy density of radiation has the form $\rho\propto
a^{-4}=e^{-4\alpha}$, which results in the Hamiltonian constraint
$C=p_+^2 - p_{\alpha}^2 + e^{2\alpha}\nu$. The constraint condition
may be implemented effectively in fashion a very similar to the way it
was done in Section~\ref{sec:eff}; however, the physical inner product
treatment would require a detailed knowledge of the (now continuous)
spectrum of an operator that is completely different from $\hat{H}$\
that was used in~(\ref{eq:schroedinger}) and is not straightforward to
obtain. Even if we could determine energy eigenstates, expanding
Gaussians or other general semiclassical states in this basis would be
challenging.  For this reason we do not attempt to implement the
constraint of the radiation-filled model on the kinematical Hilbert
space and restrict our analysis to the effective procedure,
demonstrating its wide applicability.

\subsection{Classical behavior}

\begin{center}
\begin{figure}[htbp!]
\includegraphics[width=12cm]{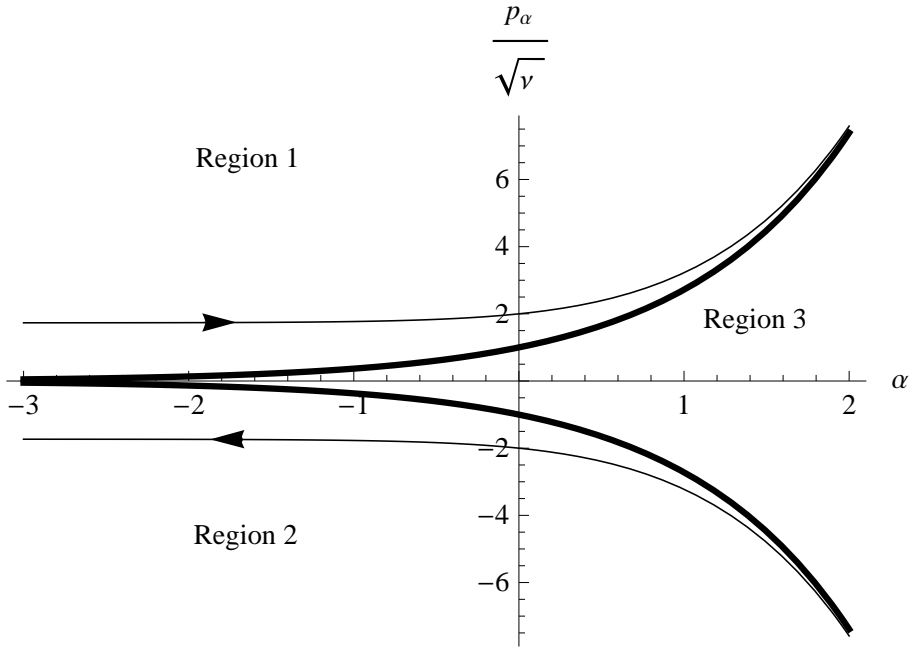}
\caption{\label{fig:20} Disjoint regions of classical solutions.
Region 1: trajectories have the shape $p_{\alpha} =
\sqrt{e^{2\alpha}\nu + {\rm const}}$, they correspond to expanding
universes as $\alpha$\ increases with $\beta_+$. Region 2:
trajectories have the shape $p_{\alpha} = -\sqrt{e^{2\alpha}\nu +
{\rm const}}$, they correspond to contracting universes as $\alpha$\
decreases with $\beta_+$. Region 3: no classical solutions.}
\end{figure}
\end{center}

We deparametrize the constraint exactly as in
Section~\ref{sec:classical}, selecting $\beta_+$\ as time. The
dynamics on $\alpha$\ and $p_{\alpha}$\ is then generated by the
Hamiltonian $H = \sqrt{p_{\alpha}^2 - e^{2\alpha}\nu}$ and results
in the equations of motion $\md\alpha/\md\beta_+= p_{\alpha}/H$ and
$\md p_{\alpha}/\md\beta_+= e^{2\alpha}\nu/H$.  Noting that $H$\ is
once again a constant of motion we can immediately infer the
classical phase-space trajectories, which are of the form
$p_{\alpha} = \pm \sqrt{e^{2\alpha}\nu + {\rm const}}$. They split
the $\alpha-p_{\alpha}$\ space into three regions as illustrated in
Fig.~\ref{fig:20}. There are no classical orbits in Region~3, as
$H$ becomes complex.
The explicit solution to the equations of motion for evolution in
terms of $\beta_+$\ is given by
\begin{equation}\label{alpharad}
\alpha (\beta_+) = \beta_+ -\log\left(2(1+Ae^{2\beta_+})\right) + B
\end{equation}
where the integration constants $A$\ and $B$\ can be related to the
initial values $\alpha_0:=\alpha(0)$\ and $p_{\alpha 0} :=
p_{\alpha}(0)$\ via
\[
A = 1 + \frac{2p_{\alpha 0}}{e^{2\alpha_0}\nu} \left(\sqrt{p_{\alpha
0}^2 - e^{2\alpha}\nu} - p_{\alpha 0} \right)
\]
\[
B = \log \left( 4 + \frac{4p_{\alpha 0}}{e^{2\alpha_0}\nu} \left(
\sqrt{p_{\alpha 0}^2 - e^{2\alpha}\nu} - p_{\alpha 0} \right)
\right) + \alpha_0
\]
In terms of $A$\ and $B$, the constant of motion is
\[
H^2 = \frac{-e^{2B}\nu}{16A}
\]
which requires $A<0$ in order for $H$\ to be real.  Using
(\ref{alpharad}), $p_{\alpha}(\beta_+)$\ may be recovered from the
equations of motion as $p_{\alpha}(\beta_+) = H
\md\alpha/\md\beta_+$. 

For the orbits in Region~1, $\alpha$\ and $p_{\alpha}$\ reach infinity
at a finite positive value of the evolution parameter, namely at
$\beta_+ = \alpha_0 - \log(p_{\alpha 0} - H)$. Explicit integration of
the expression~(\ref{eq:proper}) for the proper time using the above
solution for $\alpha(\beta_+)$\ shows that this value of $\beta_+$\ is
reached in infinite proper time. In other words, the expansion, as
expected for the radiation filled universe, takes infinite proper
time, but anisotropy asymptotically approaches a maximum value. For
the orbits in Region~2 one can obtain a similar result by tracing
evolution backwards in time; in this case $\alpha$\ reaches infinity
and $p_{\alpha}$\ reaches negative infinity at $\beta_+ = \alpha_0 -
\log(|p_{\alpha 0}| + H)$, which is always negative in that region.

Following those orbits forward in time, one finds that the collapse
happens only as $\beta_+$\ goes to infinity. Once again one can
use~(\ref{eq:proper}) to convert this to a proper time interval, with
the result that, as one would expect, the collapse takes a finite amount
of proper time,
\[
\Delta \tau := \tau(\beta_+=\infty) - \tau(\beta_+=0) =  \frac{a_0^3
e^{2B}}{32 \sqrt{|A|\nu}} \left( \frac{1+|A|}{(|A| - 1)^2} -
\frac{\cosh^{-1}\left(\sqrt{|A|}\right)}{\sqrt{|A|}}\right)\,.
\]
In this model, with positive energy, the singularity is certainly not
resolved.

\subsection{Effective constraints}

Following the effective procedure for solving constraints outlined
in Section~\ref{sec:eff} we find the constraint functions truncated
at second order:
\begin{eqnarray}
C&=& \langle\hat{p}_+\rangle^2 - \langle\hat{p}_{\alpha}\rangle^2 +
e^{2\langle\hat{\alpha}\rangle}\nu + (\Delta p_+)^2 - (\Delta
p_{\alpha})^2 +
2e^{2\langle\hat{\alpha}\rangle}\nu(\Delta \alpha)^2 = 0 \\
C_{\beta_+}& =& 2\langle\hat{p}_+\rangle \Delta(\beta_+ p_+) +
i\hbar \langle\hat{p}_+\rangle - 2\langle\hat{p}_{\alpha}\rangle
\Delta(\beta_+ p_{\alpha}) + 2e^{2\langle\hat{\alpha}\rangle}\nu
\Delta(\beta_+\alpha) = 0 \\
C_{p_+}& =& 2\langle\hat{p}_+\rangle (\Delta p_+)^2 -
2\langle\hat{p}_{\alpha}\rangle  \Delta(p_+p_{\alpha})
+ 2e^{2\langle\hat{\alpha}\rangle}\nu\Delta(p_+\alpha) = 0\\
C_{\alpha} & =& 2\langle\hat{p}_+\rangle \Delta(p_+\alpha) -
2\langle\hat{p}_{\alpha}\rangle \Delta(\alpha p_{\alpha}) - i\hbar
\langle\hat{p}_{\alpha}\rangle+
2e^{2\langle\hat{\alpha}\rangle}\nu(\Delta \alpha)^2 = 0 \\
C_{p_{\alpha}}& =& 2\langle\hat{p}_+\rangle \Delta(p_+p_{\alpha}) -
2\langle\hat{p}_{\alpha}\rangle(\Delta p_{\alpha})^2
+2e^{2\langle\hat{\alpha}\rangle}\nu \Delta(\alpha p_{\alpha})
-i\hbar e^{2\langle\hat{\alpha}\rangle}\nu= 0\,.
\label{eq:2_ord_constr_rad}
\end{eqnarray}
These can be solved and gauge fixed following the same steps as for
the model of Section~\ref{sec:eff}, with the result that evolution
in $\beta_+$\ on the expectation values and moments of
$\hat{\alpha}$\ and $\hat{p}_{\alpha}$\ is generated by the quantum
Hamiltonian
\begin{eqnarray}
H_Q &=& \sqrt{\langle\hat{p}_{\alpha}\rangle^2 -
e^{2\langle\hat{\alpha}\rangle}\nu} \Biggl( 1 +
e^{2\langle\hat{\alpha}\rangle}\nu\frac{-(\Delta p_{\alpha} )^2 + 2
\langle\hat{p}_{\alpha}\rangle \Delta(\alpha p_{\alpha}) +
(e^{2\langle\hat{\alpha}\rangle}\nu - 2
\langle\hat{p}_{\alpha}\rangle^2) (\Delta
\alpha)^2}{2(\langle\hat{p}_{\alpha}\rangle^2
- e^{2\langle\hat{\alpha}\rangle}\nu)^2} \Biggr) \nonumber \\
&&+ O\left((\Delta p_{\alpha})^4\right) + O\left((\Delta
\alpha)^4\right) + O\left((\Delta (\alpha p_{\alpha}))^2\right) \,.
\end{eqnarray}
The equations of motion are obtained by taking the quantum Poisson
bracket between quantum variables and the quantum Hamiltonian.
\begin{eqnarray}
\frac{\rm d \langle\hat{\alpha}\rangle }{{\rm d} \beta_+} &=&
\frac{\langle\hat{p}_{\alpha}\rangle}{\sqrt{\langle\hat{p}_{\alpha}\rangle^2
- e^{2\langle\hat{\alpha}\rangle}\nu}}\label{eq:eff1st2}\\  && +
e^{2\langle\hat{\alpha}\rangle}\nu\frac{\langle\hat{p}_{\alpha}\rangle
\left(\langle\hat{p}_{\alpha}\rangle ^2 +
\frac{1}{2}e^{2\langle\hat{\alpha}\rangle}\nu \right) (\Delta
\alpha)^2 -
 \Delta(\alpha p_{\alpha})
\left( 2\langle\hat{p}_{\alpha}\rangle^2 +
e^{2\langle\hat{\alpha}\rangle}\nu \right) -
\frac{3}{2}\langle\hat{p}_{\alpha}\rangle (\Delta p_{\alpha})^2}{2
\left(\langle\hat{p}_{\alpha}\rangle^2 -
e^{2\langle\hat{\alpha}\rangle}\nu\right)^{\frac{5}{2}}}\nonumber
\\\frac{\md \langle\hat{p}_{\alpha}\rangle }{{\rm d} \beta_+} &=&
\frac{-\langle\hat{\alpha}\rangle}{\sqrt{\langle\hat{p}_{\alpha}\rangle^2
- e^{2\langle\hat{\alpha}\rangle}\nu}}\\
&& + e^{2\langle\hat{\alpha}\rangle}\nu\frac{
(\frac{1}{2}e^{4\langle\hat{\alpha}\rangle}\nu^2-
\langle\hat{p}_{\alpha}\rangle^2e^{2\langle\hat{\alpha}\rangle}\nu +
2\langle\hat{p}_{\alpha}\rangle^4) (\Delta \alpha)^2  -
  2\Delta(\alpha p_{\alpha}) \left(\frac{1}{2}
e^{2\langle\hat{\alpha}\rangle}\nu +
\langle\hat{p}_{\alpha}\rangle^2\right)
\langle\hat{p}_{\alpha}\rangle}{\left(
\langle\hat{p}_{\alpha}\rangle^2
- e^{2\langle\hat{\alpha}\rangle}\nu\right)^{\frac{3}{2}}}\nonumber\\
&&
 + e^{2\langle\hat{\alpha}\rangle}\nu\frac{(\Delta p_{\alpha})^2 \left(
\langle\hat{p}_{\alpha}\rangle^2 + \frac{1}{2}
e^{2\langle\hat{\alpha}\rangle}\nu\right)}{2 \left(
\langle\hat{p}_{\alpha}\rangle^2 -
e^{2\langle\hat{\alpha}\rangle}\nu\right)^{\frac{5}{2}}} \nonumber \\
\frac{\md (\Delta\alpha)^2}{{\rm d} \beta_+} &=&
\frac{2e^{2\langle\hat{\alpha}\rangle}\nu
\left(\langle\hat{p}_{\alpha}\rangle (\Delta \alpha)^2 -
\Delta(\alpha p_{\alpha})\right)}{\left(
\langle\hat{p}_{\alpha}\rangle^2
- e^{2\langle\hat{\alpha}\rangle}\nu\right)^{\frac{3}{2}}}\\
\frac{\md (\Delta p_{\alpha})^2}{{\rm d} \beta_+} &=&
\frac{2e^{2\langle\hat{\alpha}\rangle}\nu \left( \left(
2\langle\hat{p}_{\alpha}\rangle^2 -
e^{2\langle\hat{\alpha}\rangle}\nu\right)\Delta(\alpha p_{\alpha}) -
\langle\hat{p}_{\alpha}\rangle (\Delta p_{\alpha})^2\right)}{\left(
\langle\hat{p}_{\alpha}\rangle^2 - e^{2\langle\hat{\alpha}\rangle}\nu \right)^{\frac{3}{2}}}\\
\label{eq:efflast2} \frac{\md \Delta(\alpha p_{\alpha}) }{{\rm d}
\beta_+} &=& \frac{ e^{2\langle\hat{\alpha}\rangle}\nu \left( \left(
2\langle\hat{p}_{\alpha}\rangle^2 -
e^{2\langle\hat{\alpha}\rangle}\nu\right) (\Delta \alpha)^2 -
(\Delta p_{\alpha})^2\right)}{\left(
\langle\hat{p}_{\alpha}\rangle^2 -
e^{2\langle\hat{\alpha}\rangle}\nu\right)^{\frac{3}{2}}}\,.
\end{eqnarray}
As before, these equations reduce to the classical equations of
motion at zeroth order in moments and are straightforward to evolve
numerically.

\subsection{Numerical evolution}

\begin{center}
\begin{figure}[htbp!]
\includegraphics[width=12cm]{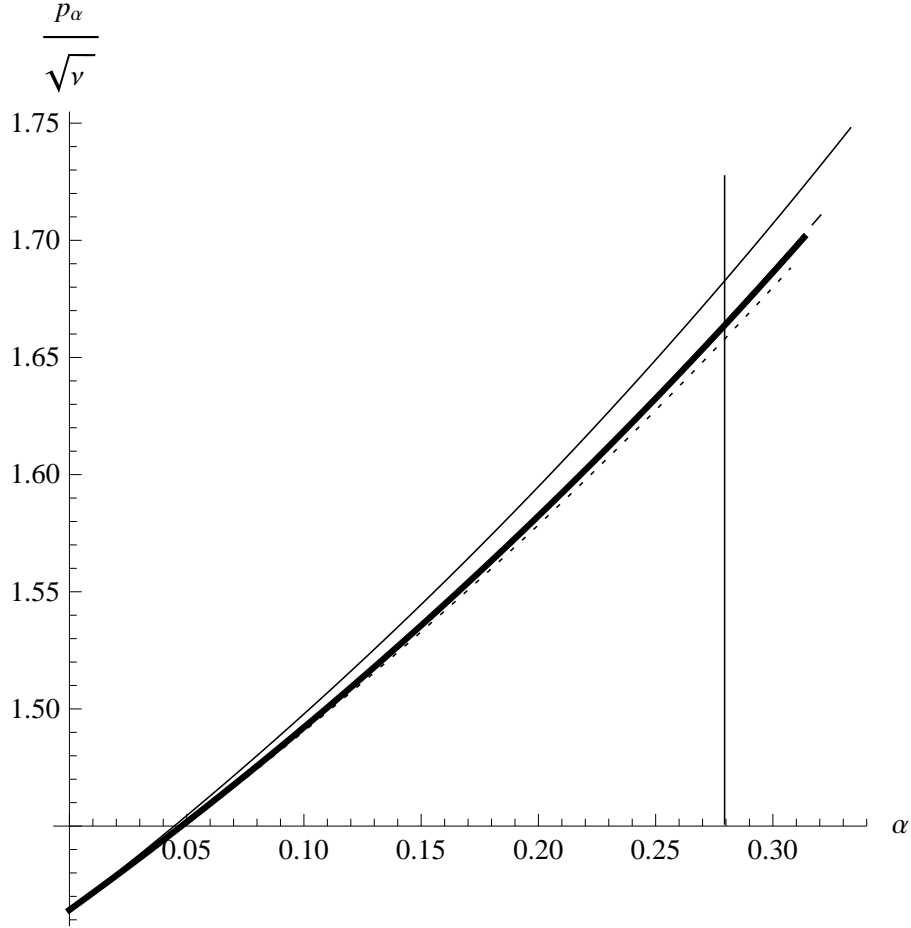}
\caption{\label{fig:21} Phase space trajectories of an expanding
universe classical (dotted) and effective with different initial
values for second order moments: dashed line---$(\Delta
\alpha)^2\nu=.001 H_0^2$, $(\Delta p_{\alpha})^2 = .025 H_0^2$,
$\Delta (\alpha p_{\alpha})\sqrt{\nu} = 0$; thin solid line---$(\Delta
\alpha)^2\nu=.025 H_0^2$, $(\Delta p_{\alpha})^2 = .001 H_0^2$,
$\Delta (\alpha p_{\alpha})\sqrt{\nu} = 0$; thick solid
line---$(\Delta \alpha)^2\nu=.008 H_0^2$, $(\Delta p_{\alpha})^2 =
.008 H_0^2$, $\Delta (\alpha p_{\alpha})\sqrt{\nu} = .005 H_0^2$.
Solutions were evolved for $0<\beta_+<.2$; the vertical line indicates
the breakdown of the semiclassical approximation based on the size of
second order moments.}
\end{figure}
\end{center}

In this section we take a classical phase space trajectory from each
of the Regions 1 and 2 and compare the effective trajectories for a
variety of semiclassical states initially peaked about these classical
solutions. Effective and classical trajectories of the expanding
universe are plotted in Fig.~\ref{fig:21}. Clearly, the significance
of quantum back-reaction can be seen well before the approximation
breaks down.

\begin{center}
\begin{figure}[htbp!]
\includegraphics[width=12cm]{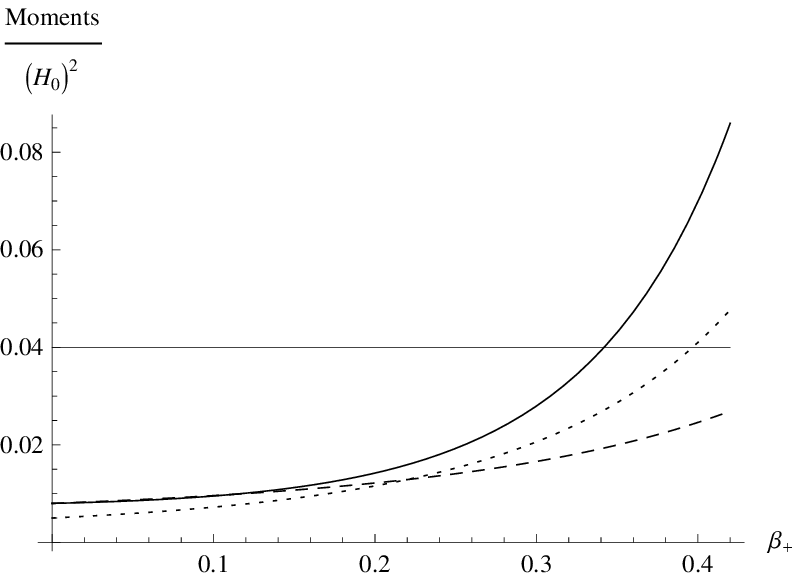}
\caption{\label{fig:22a} Evolution of second order moments in an
expanding universe. $(\Delta \alpha)^2\nu$ (dashed), $(\Delta
p_{\alpha})^2$ (solid),  $\Delta (\alpha p_{\alpha})\sqrt{\nu}$
(dotted), with initial values: $(\Delta \alpha)^2\nu=.008 H_0^2$,
$(\Delta p_{\alpha})^2 = .008 H_0^2$, $\Delta (\alpha
p_{\alpha})\sqrt{\nu} = .005 H_0^2$.}
\end{figure}
\end{center}

\begin{center}
\begin{figure}[htbp!]
\includegraphics[width=12cm]{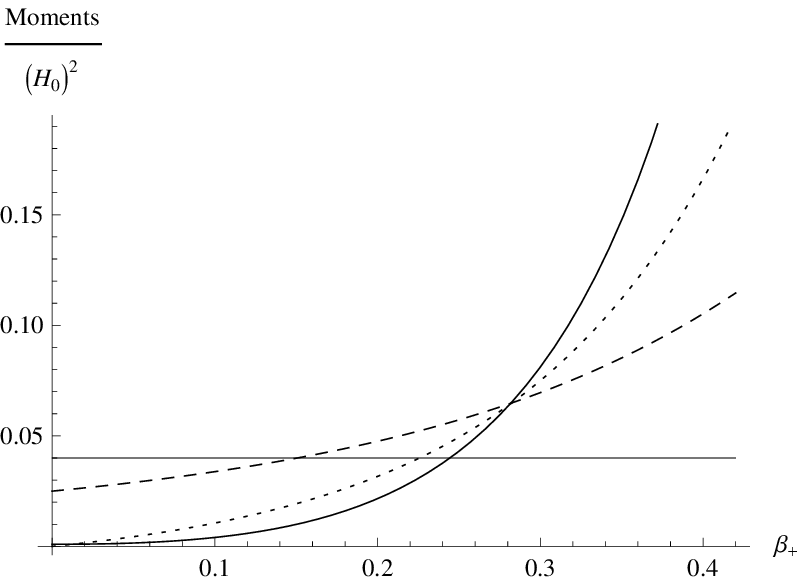}
\caption{\label{fig:22b} Evolution of second order moments in an
expanding universe. $(\Delta \alpha)^2\nu$ (dashed), $(\Delta
p_{\alpha})^2$ (solid),  $\Delta (\alpha p_{\alpha})\sqrt{\nu}$
(dotted), with initial values: $(\Delta \alpha)^2\nu=.025 H_0^2$,
$(\Delta p_{\alpha})^2 = .001 H_0^2$, $\Delta (\alpha
p_{\alpha})\sqrt{\nu} = 0$.}
\end{figure}
\end{center}

\begin{center}
\begin{figure}[htbp!]
\includegraphics[width=12cm]{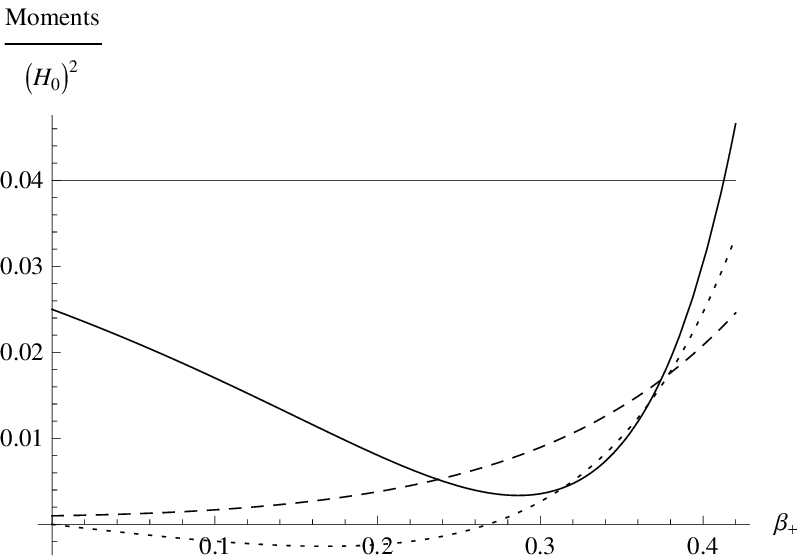}
\caption{\label{fig:22c} Evolution of second order moments in an
expanding universe. $(\Delta \alpha)^2\nu$ (dashed), $(\Delta
p_{\alpha})^2$ (solid),  $\Delta (\alpha p_{\alpha})\sqrt{\nu}$
(dotted), with initial values: $(\Delta \alpha)^2\nu=.001 H_0^2$,
$(\Delta p_{\alpha})^2 = .025 H_0^2$, $\Delta (\alpha
p_{\alpha})\sqrt{\nu} = 0$.}
\end{figure}
\end{center}

The corresponding evolution of the leading order moments starting
from different initial values is plotted in
Figs.~\ref{fig:22a}-\ref{fig:22c}, where the horizontal line, as
before, indicates an approximate threshold of the semiclassical
approximation for the second order moments.

\begin{center}
\begin{figure}[htbp!]
\includegraphics[width=7.5cm]{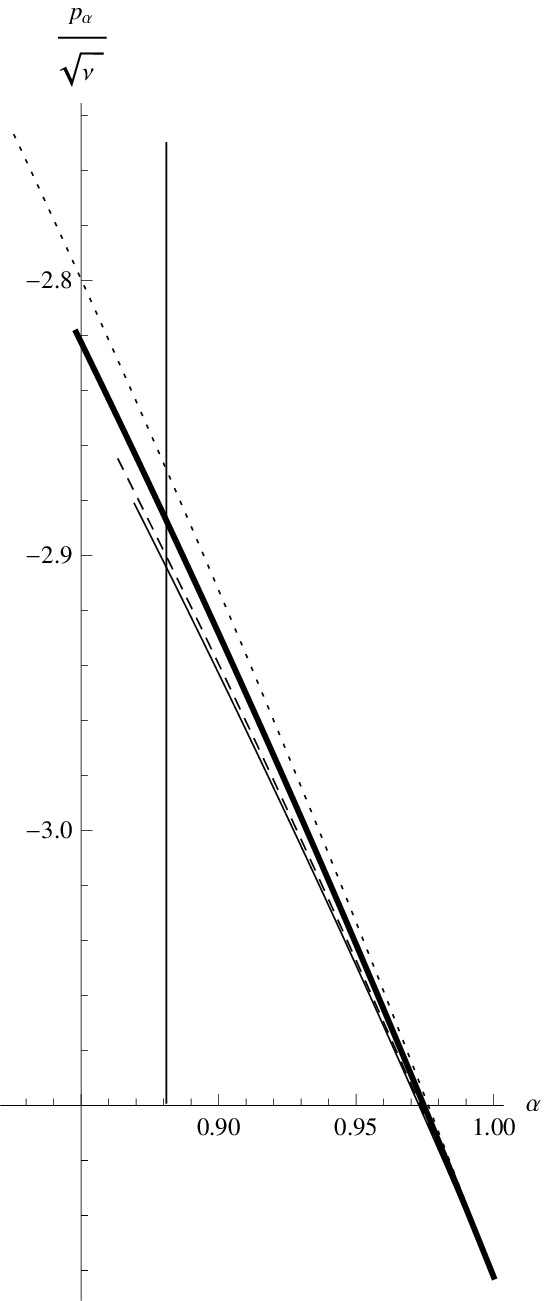}
\caption{\label{fig:23} Phase space trajectories of a contracting
universe classical (thin solid line) and effective with different
initial values for second order moments: dashed line---$(\Delta
\alpha)^2\nu=.001 H_0^2$, $(\Delta p_{\alpha})^2 = .025 H_0^2$,
$\Delta (\alpha p_{\alpha})\sqrt{\nu} = 0$; dotted line---$(\Delta
\alpha)^2\nu=.025 H_0^2$, $(\Delta p_{\alpha})^2 = .001 H_0^2$,
$\Delta (\alpha p_{\alpha})\sqrt{\nu} = 0$; thick solid
line---$(\Delta \alpha)^2\nu=.008 H_0^2$, $(\Delta p_{\alpha})^2 =
.008 H_0^2$, $\Delta (\alpha p_{\alpha})\sqrt{\nu} = .005 H_0^2$.
Solutions were evolved for $0<\beta_+<.07$; the vertical line
indicates the breakdown of the semiclassical approximation (coming
from the right along the $\alpha$-axis).}
\end{figure}
\end{center}

\begin{center}
\begin{figure}[htbp!]
\includegraphics[width=12cm]{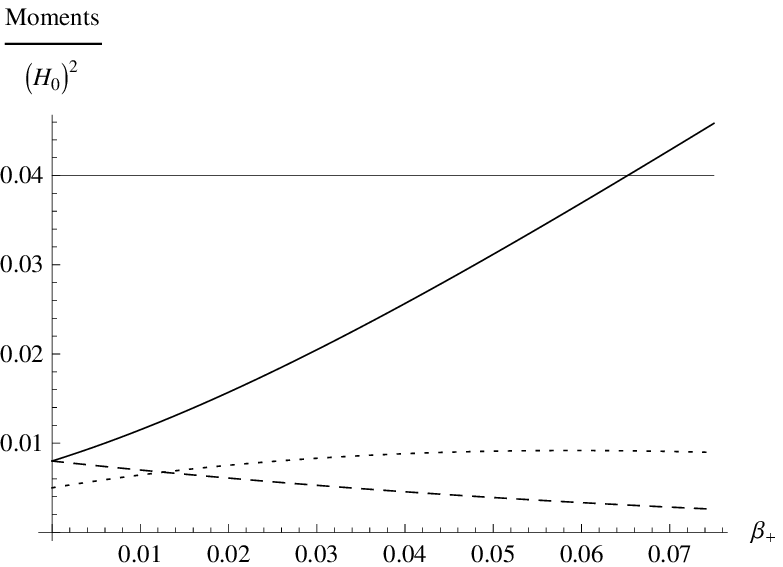}
\caption{\label{fig:24a} Evolution of second order moments in a
contracting universe. $(\Delta \alpha)^2\nu$ (dashed), $(\Delta
p_{\alpha})^2$ (solid),  $\Delta (\alpha p_{\alpha})\sqrt{\nu}$
(dotted), with initial values: $(\Delta \alpha)^2\nu=.008 H_0^2$,
$(\Delta p_{\alpha})^2 = .008 H_0^2$, $\Delta (\alpha
p_{\alpha})\sqrt{\nu} = .005 H_0^2$.}
\end{figure}
\end{center}

\begin{center}
\begin{figure}[htbp!]
\includegraphics[width=12cm]{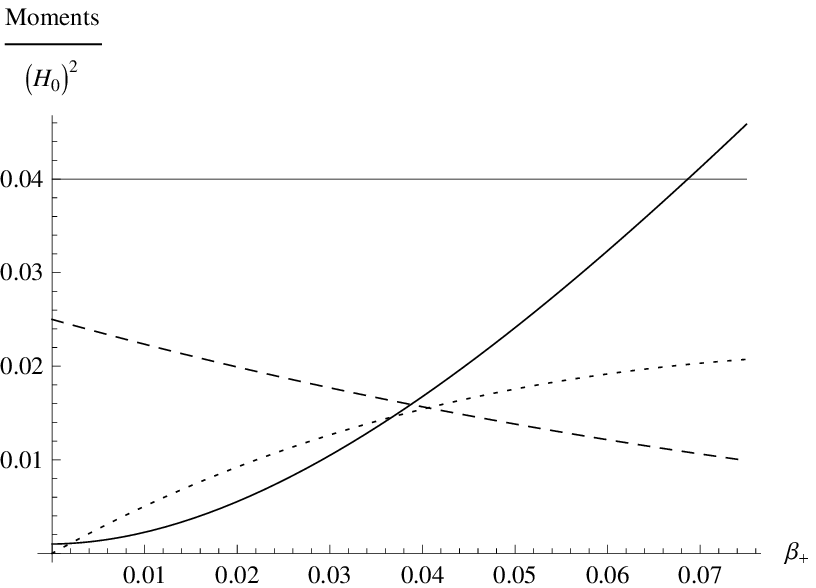}
\caption{\label{fig:24b} Evolution of second order moments in a
contracting universe. $(\Delta \alpha)^2\nu$ (dashed), $(\Delta
p_{\alpha})^2$ (solid),  $\Delta (\alpha p_{\alpha})\sqrt{\nu}$
(dotted), with initial values: $(\Delta \alpha)^2\nu=.025 H_0^2$,
$(\Delta p_{\alpha})^2 = .001 H_0^2$, $\Delta (\alpha
p_{\alpha})\sqrt{\nu} = 0$.}
\end{figure}
\end{center}

\begin{center}
\begin{figure}[htbp!]
\includegraphics[width=12cm]{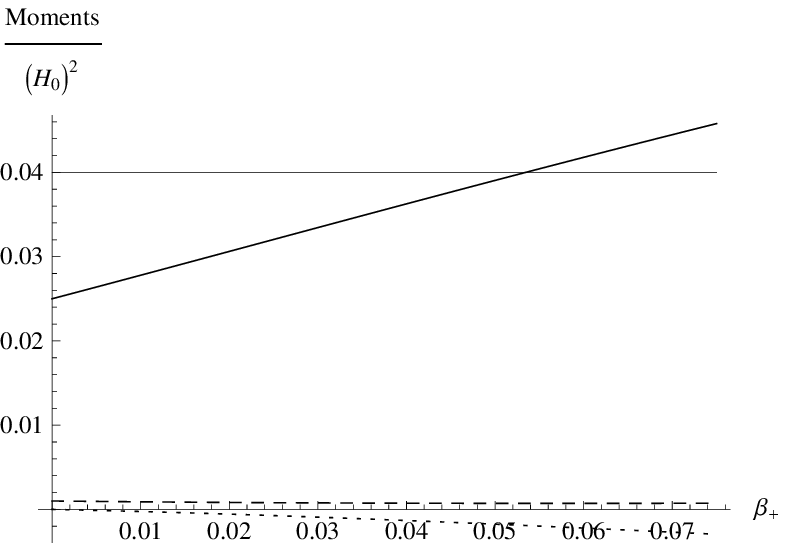}
\caption{\label{fig:24c} Evolution of second order moments in a
contracting universe. $(\Delta \alpha)^2\nu$ (dashed), $(\Delta
p_{\alpha})^2$ (solid),  $\Delta (\alpha p_{\alpha})\sqrt{\nu}$
(dotted), with initial values: $(\Delta \alpha)^2\nu=.001 H_0^2$,
$(\Delta p_{\alpha})^2 = .025 H_0^2$, $\Delta (\alpha
p_{\alpha})\sqrt{\nu} = 0$.}
\end{figure}
\end{center}

Effective and classical trajectories of the contracting universe are
plotted in Fig.~\ref{fig:23}.
The corresponding evolution of the leading order moments starting
from different initial values is plotted in
Figs.~\ref{fig:24a}-\ref{fig:24c}, where the horizontal line, once
again, indicates an approximate threshold of the semiclassical
approximation for the second order moments.

\section{Conclusions}

Most quantum systems of physical relevance can only be analyzed by
perturbation methods. Quantum gravity and quantum cosmology cannot be
considered exceptions. As demonstrated by the examples of this
article, canonical effective equations, based on the back-reaction of
moments of a state on its expectation values, are of wide
applicability, capture quantum effects reliably, and approximate the
full quantum dynamics in a self-consistent way. They are tractable
even in situations where semiclassical wave functions in physical Hilbert
spaces would be too complicated to be constructed --- cases which
abound in quantum cosmology and quantum gravity.

\section*{Acknowledgements}

This work was supported in part by NSF grant PHY 0748336.


\end{document}